\documentclass[journal,12pt,onecolumn,draftclsnofoot]{IEEEtran}

\usepackage[letterpaper, left=1.01in, right=1.01in, bottom=1in, top=0.75in]{geometry}

\usepackage[utf8]{inputenc}
\usepackage[pdftex,final]{graphicx}
\usepackage{amsmath}
\usepackage{amsthm}
\usepackage{mathrsfs}
\usepackage{lscape}
\usepackage{latexsym}
\usepackage{amssymb}
\usepackage{bm}
\usepackage{bbm}
\usepackage{cite}
\usepackage{url}
\usepackage{xcolor}
\usepackage[caption=false,font=footnotesize]{subfig}
\usepackage{lmodern}
\usepackage{tikz}
\usepackage{tikzsymbols}
\usepackage{float}
\usepackage{circuitikz}
\usepackage[colorinlistoftodos,textwidth=\marginparwidth]{todonotes}
\usetikzlibrary{shapes.geometric, arrows, positioning, fit}

\newsavebox\dotbox
\sbox{\dotbox}{\(\displaystyle\bigodot\)}

\newtheorem{remark}{Remark}

\title{Bandwidth-Agile Image Transmission \\with Deep Joint Source-Channel Coding} 
\author{\IEEEauthorblockN{David Burth Kurka  and  Deniz G\"und\"uz}

\IEEEauthorblockA{\small Information Processing and Communications Laboratory\\
Department of Electrical and Electronic Engineering\\
Imperial College London, London, UK  \\
{\tt \{d.kurka, d.gunduz\}@imperial.ac.uk}
}
}
\date{}

\begin{document}

\maketitle

\begin{abstract}

We propose deep learning based communication methods for adaptive-bandwidth transmission of images over wireless channels.
We consider the scenario in which images are transmitted progressively in layers over time or frequency, and such layers can be aggregated by receivers in order to increase the quality of their reconstructions.
We investigate two scenarios, one in which the layers are sent sequentially, and incrementally contribute to the refinement of a reconstruction, and another in which the layers are independent and can be retrieved in any order. Those scenarios correspond to the well known problems of \textit{successive refinement} and \textit{multiple descriptions}, respectively, in the context of joint source-channel coding (JSCC).
We propose DeepJSCC-$l$, an innovative solution that uses convolutional autoencoders, and present three architectures with different complexity trade-offs. To the best of our knowledge, this is the first practical multiple-description JSCC scheme developed and tested for practical information sources and channels.
Numerical results show that DeepJSCC-$l$ can learn to transmit the source progressively with negligible losses in the end-to-end performance compared with a single transmission. Moreover, DeepJSCC-$l$ has comparable performance with state of the art digital progressive transmission schemes in the challenging low signal-to-noise ratio (SNR) and small bandwidth regimes, with the additional advantage of graceful degradation with channel SNR.
\end{abstract}


\section{Introduction}\label{s:Intro}

We consider wireless transmission of images in multiple layers, each communicated over an independent noisy channel. The receiver receives the output of only a subset of the channels, and tries to reconstruct the original image at the best quality possible. We would like the image quality to increase as more layers are received. Such a scheme enables flexible transmission modes, where communication can be fulfilled with varying bandwidth availability. For example, these layers may be communicated over different frequency bands, and the receiver may be able to tune into only a subset of these bands. We would like the receiver to be able to reconstruct the underlying image no matter which subset of bands it can tune into. Alternatively, if the layers are transmitted sequentially in time, the receiver can stop receiving if it has reached a desired reconstruction quality, saving valuable time and energy resources. Concurrently, other receivers may continue to receive more layers, and can recover a better quality reconstruction by receiving additional symbols. 
Such a scheme results in bandwidth agile communication, and can be used in a variety of applications in which communication is either expensive, urgent, or limited. For example, in surveillance applications, it may be beneficial to quickly send a low-resolution image to detect a potential threat as soon as possible, while a higher resolution description can be received with additional delay for further evaluation or archival purposes. This approach can also benefit emergency systems, where urgent actions may need to be taken based on low resolution signals transmitted rapidly.

\begin{figure*}[t]
	\begin{center}
   \includegraphics[width=.9\linewidth]{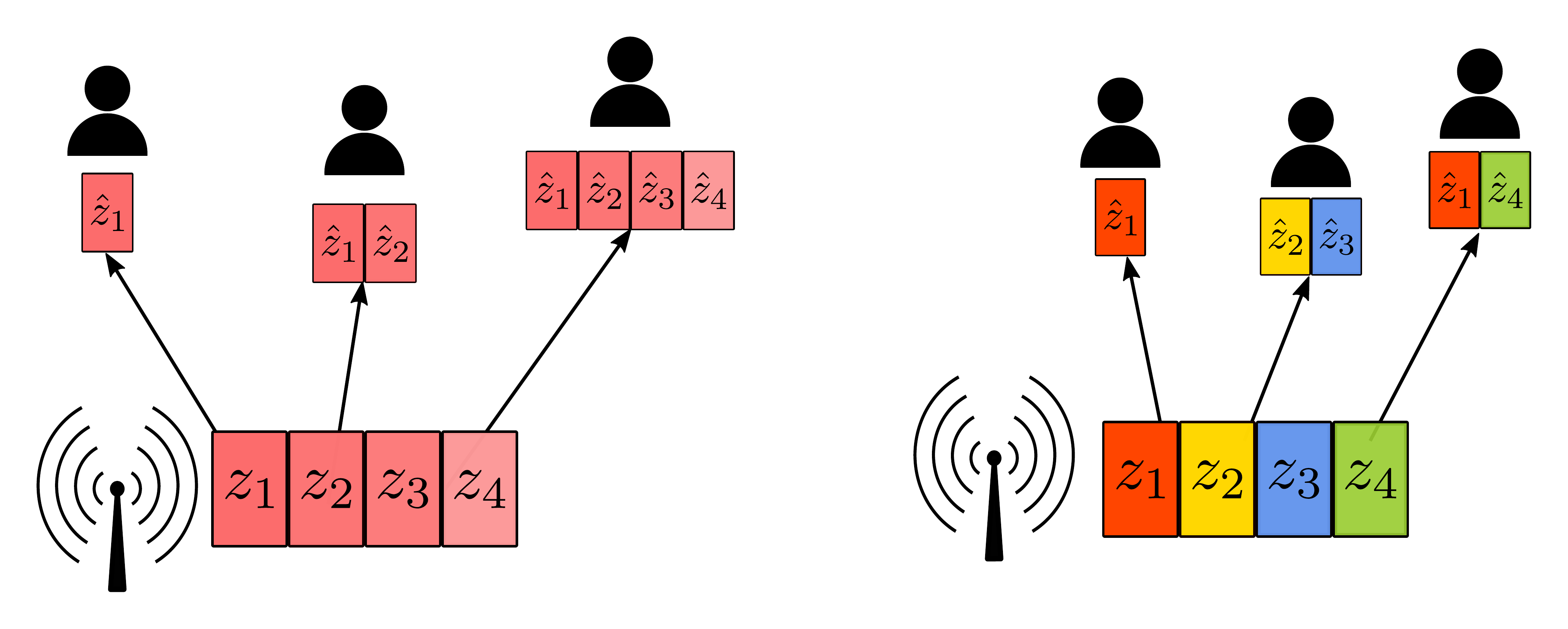}
	\end{center}
  \caption{Bandwidth-agile JSCC illustrating successive refinement (left) and multiple descriptions (right). Given an input image, the transmitter generates multiple codewords, denoted by $\bm{z}_i$, $i \in 1, \dots, L$ (in the illustration, $L=4$), and the receiver receives noisy versions of a subset of the codewords. Under the successive refinement scheme, layers are received sequentially, i.e., $\bm{\hat z}_1, \dots, \bm{\hat z}_i$ for $i \leq L$. Under the more general multiple description scenario, any subset of noisy codewords may be received.}
  \label{fig:modelillustration}
\end{figure*}

Our problem is the joint source-channel coding (JSCC) version of the well-known \textit{multiple descriptions} problem \cite{Goyal:SPM:01}. The conventional multiple description problem focuses on the compression aspects, where the image is compressed into multiple layers, each at a different rate. In the multiple description problem, each layer is either received perfectly or not received at all, and the goal is to obtain the best possible reconstruction quality for any subset of received layers. A special case of this problem is the \textit{successive refinement} problem, in which the layers are transmitted sequentially, starting from a base layer providing the main elements of the content being transmitted, followed by refinement layers used to enhance the image quality and add details to it. See Fig.~\ref{fig:modelillustration} for an illustration of the two problems.

The rate-distortion region for both the multiple description \cite{gray_source_1974, Wolf:MultDescr:1980, ElGamal:MultDescr:1982} and the successive refinement problems \cite{koshelev1980hierarchical, Equitz1991, Rimoldi1994, Nayak:IT:10} have been studied extensively from an information theoretic perspective. While the optimal rate-distortion region for the multiple description problem remains open for general source distributions, optimal characterization is known for Gaussian sources \cite{Ozarow1980source}. A general single-letter characterization of the rate-distortion region is possible for the successive refinement problem \cite{Rimoldi1994}. Generating practical multiple description and successive refinement codes has also been studied. While the best practical source codes typically depend on the statistical properties of the underlying source distribution, researchers have studied how to achieve successive refinement or multiple descriptions through quantization \cite{Vaishampayan:IT:93, Vaishampayan:IT:01, Dayan:IT:02, Jafarkhani:TC:99}. Multiple descriptions can also be obtained through a pair of correlating transforms \cite{Wang:TIP:01}.

The JSCC version of the problem, however, has received considerably less attention. This may be partially due to the theoretical optimality of separation between the source and channel coding problems. A separation theorem is proven in \cite{Steinberg:IT:06} for the successive refinement JSCC problem when the layers are transmitted over independent channels. It is shown that it is optimal to compress the source into multiple layers using successive refinement source coding, where the rate of each layer is dictated by the capacity of the channel it is transmitted over. A similar result is proven for the multiple description problem for Gaussian sources in \cite{Gastpar:PhD}. Note, however, that the optimality of separation holds only under the information theoretic assumption of ergodic sources and channels, and in the asymptotic limits of large source and channel blocklengths and unbounded complexity.

Here, following our previous work~\cite{bourtsoulatze:TCCN:2019,bourtsoulatze:ICASSP:2019,kurka:IZS:2020}, we use deep learning (DL) methods, in particular, the autoencoder architecture \cite{GoodfellowDL2016}, for the design of a practical end-to-end multiple description image transmission system.
In~\cite{bourtsoulatze:TCCN:2019}, we introduced a novel end-to-end DL-based JSCC scheme for image transmission over wireless channels, called \emph{DeepJSCC}, where encoding and decoding functions are parameterized by convolutional neural networks (CNNs) and the communication channel is incorporated into the neural network (NN) architecture as a non-trainable layer. This method achieves remarkable performance in low signal-to-noise ratio (SNR) and limited channel bandwidth, also showing resilience to mismatch between training and test channel conditions and channel variations, similarly to analog communications; that is, it does not suffer from the `cliff effect' unlike digital communication schemes based on separate source and channel coding. JSCC of text has also been studied in \cite{FarsadICASSP2018}. Several recent works have considered variational autoencoders for JSCC \cite{gunduz:patent, Choi:arxiv:18, Saidutta:DCC:19, Saidutta:ISIT:19}. Similarly to \cite{Saidutta:DCC:19} and \cite{Saidutta:ISIT:19}, Gaussian sources are considered in \cite{Xuan:SPCOM:20}, but LSTM based autoencoder architecture is employed instead.  Extension of \cite{bourtsoulatze:TCCN:2019} to a network of orthogonal links is considered in \cite{Liu:ICC:20} with the focus on `network coding' carried out by the intermediate nodes. Techniques and ideas from DeepJSCC have also been exploited for channel state information feedback \cite{Mashhadi:ICASSP:20}, classification at the network edge \cite{Shao:ICC:20}, or wireless image retrieval \cite{Jankowski:JSAC:21}.

In parallel, there have been significant efforts in the design of DL-based image compression schemes, in some cases outperforming current handcrafted codecs \cite{TodericiICLR2016,RippelICML2017,BalleICLR2018,Balle:NIPS2018,minnen2020channelwise}. More recently, these efforts have also been extended to the multiple description problem \cite{Zhao:DCC:19, Zhao:TCS:19, Lu:Access:19}. In the source coding domain, an autoencoder is used for dimensionality reduction to efficiently represent the original source image. This is followed by quantization and entropy coding as in standard compression codecs. However, in the JSCC problem, a low dimensional representation of the source is not sufficient. The encoder must learn how to map the input to the transmitted channel input vectors. In principle, this transformation should map similar source signals to similar channel inputs, so that they can be reconstructed with minimal distortion despite channel noise.

We first tackle the successive refinement problem, and introduce a new strategy for progressive image transmission, called \emph{DeepJSCC-$l$}. We show with extensive experimental results that DeepJSCC-$l$ can successfully learn to encode images into multiple channel codewords, each one successively refining the reconstruction at the receiver, and that the introduction of multiple codewords does not cause significant performance losses. In the context of source coding, a source is said to be ``successively refinable'' under a specified distortion measure when it is possible to achieve the single layer rate distortion performance at every stage of the successive refinement process. For example, Gaussian sources are successively refinable under squared-error distortion. Here, in the context of JSCC, our experimental results suggest that natural images transmitted with DeepJSCC are nearly `successively refinable' over Gaussian channels.
We also demonstrate how the problem of successive refinement can be approached with different implementations, by proposing three candidate solutions with different time-space complexity trade-offs. Finally, we further extend the solution and explore the more general multiple description problem, showing that our solution is able to learn independent codewords that have similar performance to a single layer transmission when sent separately, yet significantly improving the transmission performance when multiple parts are combined.

Despite the introduction of progressive transmission through successive refinement, all the properties present in single-layer transmission with DeepJSCC \cite{bourtsoulatze:TCCN:2019}, such as graceful degradation, versatility in different channel models, and better or comparable performance compared to separate source and channel coding (JPEG2000 or BPG followed by high performance channel codes) are maintained. Thus, this work introduces, to the best of our knowledge, not only the first practical progressive and multiple-decription JSCC schemes for realistic information sources and channels, but also a solution that enables flexible and high-performance communication with adaptive bandwidth and uncertain channel quality; providing one more reason to explore its practical implementation in future communication systems.

In summary, the main contributions of this work are:
\begin{itemize}
    \item The first practical scheme for the successive refinement and multiple description JSCC problems, achieved by a data-driven machine-learning approach;
    \item Introduction of a family of network architectures that are able to learn solutions with different complexity trade-offs;
    \item Outstanding performance at the task of image transmission when compared to digital schemes and negligible performance compromise due to multi-channel adaptation;
    \item Adaptability to different communication channel models (AWGN, Rayleigh fading), presenting graceful degradation over non-ergodic channels.
\end{itemize}

\section{System Model}

We consider wireless transmission of images over $L$ parallel channels. Let ${\bm z}_i \in \mathbb{C}^{k_i}$ denote the complex channel input vector and $\bm {\hat z}_i \in \mathbb{C}^{k_i}$ the corresponding channel output vector for the $i$-th channel, $i \in [L]  \triangleq [1, \dots, L]$.
We assume that the transmissions of ${\bm z}_i$ sequences are done through independent realizations of a noisy communication channel represented by the transfer function $\bm {\hat z}_i = \eta({\bm z}_i)$, and consider in this work two widely used channel models: (a) the additive white Gaussian noise (AWGN) channel, and (b) the slow fading channel. The transfer function of the Gaussian channel is $\eta_{n} (\bm z_i)= \bm z_i + \bm n$, where the vector $\bm n \in \mathbb{C}^{k_i}$ consists of independent identically distributed (i.i.d.) samples from a circularly symmetric complex Gaussian distribution, i.e., $ \bm n \sim  \mathcal{CN}(0,\sigma^2\bm I_{k_{i}})$, where $\sigma^2$ is the average noise power. In the case of a slow fading channel, we adopt the commonly used Rayleigh slow fading model. The multiplicative effect of the channel gain is captured by the channel transfer function $\eta_{h}(\bm z_i) =  h \bm z_i$, where $ h \sim \mathcal{CN}(0,H_c)$ is  a complex normal random variable.

Let $\bm x \in  \mathbb{R}^n$ denote the image to be transmitted. The receiver obtains a subset $\mathscr{S} \subseteq [L]$ of the channel output vectors, and creates a reconstruction $\bm {\hat x}_\mathscr{S} \in \mathbb{R}^n$. We consider two kinds of subsets: in the \textit{successive refinement} problem, the receiver obtains channel output vectors corresponding to sequential and consecutive channels, i.e., $\mathscr{S} = [i]$ for some $1 \leq i \leq L$; in the \textit{multiple description} problem, the receiver obtains channel output vectors from arbitrary combinations of channels.
As different channel output subsets have different sizes, we achieve agile bandwidth in the sense that the same image can be transmitted and reconstructed with the use of different amounts of bandwidth.

We will call the image dimension $n$ as the \textit{source bandwidth}, and the dimension $k_i$ of the $i$-th channel as the \textit{channel bandwidth}. We will refer to the ratio $k_i/n$ as the \textit{bandwidth ratio} for the $i$-th channel.

An average power constraint is imposed on the transmitted signal at every channel, i.e,
\begin{equation}
\frac{1}{k_i} \mathbb{E}[{\bm z_i}^*{\bm z_i}] \leq P,\qquad \forall i \in [L],
\label{power_norm}    
\end{equation}

where the average signal-to-noise ratio (SNR) given by: $\mathrm{SNR} = 10\log_{10}\frac{P}{\sigma^2} ~ (dB)$.

Performance is evaluated by the peak signal to noise ratio ($\mathrm{PSNR}_j$) between the input image  $\bm x$ and a reconstruction $\bm{\hat x}_j$. The PSNR is inversely proportional to the mean square error (MSE), and both are defined as:

\begin{equation}
\mathrm{MSE}_\mathscr{S} = \frac{1}{n} ||\bm x-\bm{\hat x}_\mathscr{S} ||^2
\end{equation}
\begin{equation}
  \mathrm{PSNR}_\mathscr{S} = 10\log_{10}\frac{\mathrm{MAX}^2}{\mathrm{MSE}_\mathscr{S}},
\end{equation}
where $\mathrm{MAX}$ is the maximum value a pixel can take, which is $255$ in our case (we consider RGB images, with 8 bits per pixel per color channel).

\section{DeepJSCC-$l$}

\begin{figure}[t]
    \centering
    \includegraphics[width=0.75\linewidth]{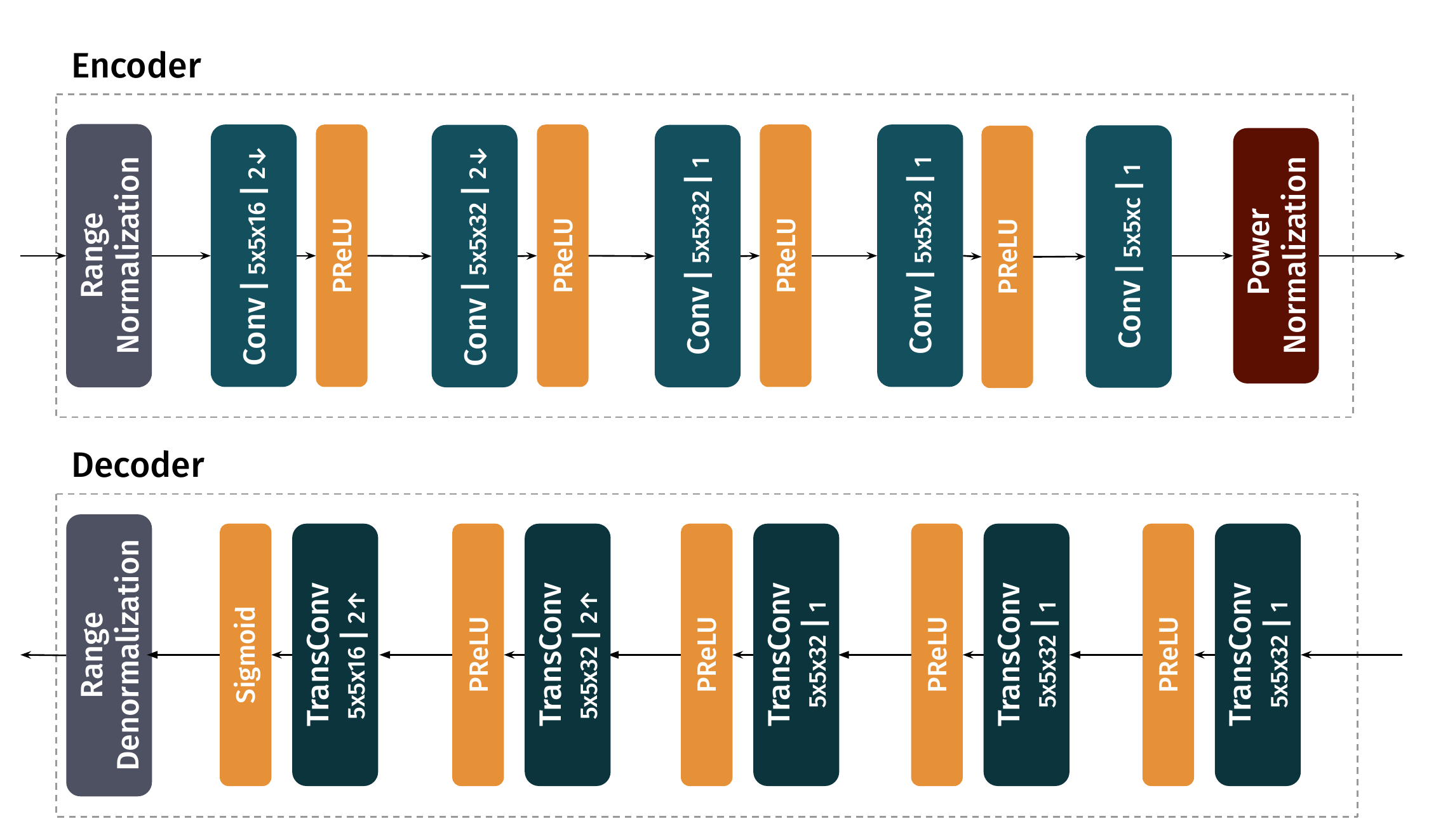}
    \caption{The encoder and decoder components used in this paper, introduced in \cite{bourtsoulatze:TCCN:2019}. The notation $k\times k \times d / s$ refers to kernel size $k$, depth $d$ and stride $s$, while $c$ defines the encoder's compression rate.}
    \label{fig:architecture}
\end{figure}

Inspired by the success of \cite{bourtsoulatze:TCCN:2019}, we propose the use of CNNs to represent both JSCC encoder and decoder, and add the channel to the model as a differentiable yet non-trainable layer, producing random values at every realization. All neural network components are trained jointly and the performance is optimized on realizations of an end-to-end communication system, forming an autoencoder architecture.
Thus, our proposed architecture, called DeepJSCC-$l$, is built using neural encoders and decoders as basic components. We will primarily investigate the model with one encoder and multiple decoders (as can be seen in Fig.~\ref{fig:modeldirectpass} for the case of successive refinement with $L=2$), but alternative models are also considered. The name DeepJSCC-$l$ refers to the family of all the different architectures and solutions considered.

The encoder is a CNN and is represented by the deterministic function $f^{\boldsymbol{\theta}}$ parameterized by vector $\boldsymbol{\theta}$. It receives as input the source image $\bm x$, and outputs at once all the channel input symbols $\bm z$, i.e., we have $\bm z = f^{\boldsymbol{\theta}}(\bm x)$ with $\bm  z \in \mathbb{C}^{k}$, where $k$ is the total bandwidth, i.e., $k = \sum_{i=1}^L k_i$. The channel input is $\bm z = (\bm z_1, \dots, \bm z_L)$, where $\bm z_i $ is transmitted over the $i$-th channel.

We consider that, for each valid subset $\mathscr{S}$ of channel output vectors received, a different decoder is employed to transform the noisy symbols into reconstruction $\bm{\hat x_{\mathscr{S}}}$. Thus, the decoder is a CNN represented by $g_\mathscr{S}^{\boldsymbol{\phi}_\mathscr{S}}$, where $\boldsymbol{\phi}_\mathscr{S}$ is the learned parameter vector. We denote the concatenation of all channel outputs for subset $\mathscr{S}$ by $\bm{\hat Z}_\mathscr{S}$, i.e., $\bm{\hat Z}_\mathscr{S} = \bigcup_{i \in \mathscr{S}} \bm{\hat z}_i$. The corresponding reconstruction is given by $\bm{\hat{x}}_\mathscr{S} = g_\mathscr{S}^{\boldsymbol{\phi}_\mathscr{S}}(\bm{\hat Z}_\mathscr{S})$.

We optimize all the parameters jointly to minimize the average distortion between the input image $\bm{x}$ and a partial reconstructions $\bm{\hat x}_\mathscr{S}$:
\begin{equation}
    (\boldsymbol{\theta}^*, \boldsymbol{\phi}^*) = \sum_{\mathscr{S}} \underset{\theta, \phi_\mathscr{S}}{\mathrm{arg\,min}}~ \mathbb{E}_{p(\bm{x},\bm{\hat{x}}_\mathscr{S})} [d(\bm{x}, \bm{\hat{x}}_\mathscr{S})],
    \label{eq:expdistortion}
\end{equation}
where $\boldsymbol{\phi}$ is the collection of of all decoders' parameter vectors, $d(\bm{x}, \bm{\hat{x}}_\mathscr{S})$ is a given distortion measure, and $p(\bm{x},\bm{\hat{x}}_\mathscr{S})$ the joint probability distribution of the original and reconstructed images, which depends on the channel and input image statistics, as well on the encoder and decoder parameters.
Note that this is a multi-objective problem, as multiple reconstructions are considered and the parameter vector $\boldsymbol{\theta}$ is common in the optimization of every reconstruction. We address this problem by performing a joint training the sum of all the objectives, or by greedily training them sequentially (Section~\ref{sec:resnet}).

Fig.~\ref{fig:architecture} presents the NN architectures used for the encoder and decoder components. The encoder and decoder are symmetric, containing the same number of convolutional layers and trainable weights. The convolutional layers are responsible for feature extraction and downsampling (at the encoder) or upsampling (at the decoder) through stride and varying the depth of the output space. The last layer of the encoder is parameterized by depth $c$, which defines the total bandwidth ratio of all the layers combined: $k/n = (H/4 \times W/4 \times c)/(H \times W \times 3) = c/48$, where $H$ and $W$ are the height and width of an image with 3 color channels. After each convolution, we use the parametric ReLU (PReLU)~\cite{PReLU} activation function, or a sigmoid in the last block of the decoder to produce outputs in the range $[0,1]$. Normalization at the beginning of the encoder, and denormalization at the end of the decoder convert values from range $[0,255]$ to $[0,1]$ (and vice versa). At the end of the encoder, the output of the last convolution layer is normalized as in Eq. (\ref{power_norm}) so the average power of each layer is constrained to $P=1$.
Note that the total channel bandwidth $k$ used by DeepJSCC-$l$ depends on the input dimension $n$; that is, the same model can output different channel bandwidths for different input sizes, keeping the bandwidth ratio constant. Thus, we will consider the bandwidth ratio ($k/n$) when presenting and comparing results.

The architecture in Fig.~\ref{fig:architecture} is based on~\cite{bourtsoulatze:TCCN:2019}, where it is employed for single-layer transmission, and is shown to achieve competitive results with the state-of-the-art separation-based  digital schemes. While we studied improved and slightly more complex versions of this architecture in a separate work \cite{kurka:IZS:2020}, we use the original DeepJSCC architecture for most of the simulations in this paper, as our focus is to show that the single-layer model can be extended to multi-layer transmission. A more robust architecture is presented in Section \ref{sec:digital}.

All simulations are implemented on TensorFlow \cite{abadi2016tensorflow}, using the Adam algorithm \cite{AdamICLR2015} for stochastic gradient descent, learning rate of $10^{-4}$, and $64$ images per training batch.
As the model is fully convolutional, a trained model can accept images of arbitrary dimensions, but the results presented in this paper are for models trained and evaluated on the CIFAR-10 dataset \cite{CIFARdataset} containing 50000 training images and 10000 test images with dimension $n = 32 \times 32 \times 3$.
Results are presented in terms of the average PSNR calculated over the whole CIFAR-10 test dataset, each image transmitted over $10$ independent realizations of the noisy channel.

\section{Successive Refinement JSCC}
\label{sec:succrefinement}

\begin{figure}
	\begin{center}
    \resizebox {0.85\linewidth} {!}%
    {

 \tikzstyle{txt} = [text centered]
 \tikzstyle{box} = [rectangle, rounded corners, minimum width=2.5cm, minimum height=1.5cm, text width=2.5cm, text centered, draw=black]
 \tikzstyle{bbox} = [rectangle, thick, minimum width=1.5cm, minimum height=1cm, text width=1.5cm, text centered, draw=black]
 \tikzstyle{arrow} = [thick,->,>=stealth]
 \tikzstyle{fitted} = [draw=gray, thick, dotted, inner sep=0.75em]


\begin{tikzpicture}[node distance=2.3cm]

\node (x) [txt, font=\fontsize{14}{0}\selectfont] {$\bm{x}$};
\node (encoder) [box, right of=x, font=\fontsize{12}{12}\selectfont] {Encoder \\ ($f_{\bm{\theta}}$)};
\node (z) [txt, right of=encoder, xshift=0.5cm, font=\fontsize{14}{0}\selectfont] { };
\node (channel1) [bbox, above right of=z,  font=\fontsize{12}{12}\selectfont, xshift=0.5cm, yshift=-0.5cm] {channel};
\node (decoder1) [box, right of=channel1, xshift=1.5cm,  font=\fontsize{12}{12}\selectfont] {Decoder1\\($g_{\bm{\phi_1}}$)};
\node (xhat1) [txt, right of=decoder1,font=\fontsize{14}{0}\selectfont] {$\bm{\hat{x}_1}$};
\node (baselayer) [fitted, fit=(channel1) (decoder1)] {};
\node at (baselayer.north) [above, inner sep=1mm] {Base Layer};

\node (channel2) [bbox, below right of=z,  font=\fontsize{12}{12}\selectfont, xshift=0.5cm, yshift=0.5cm] {channel};
\node (decoder2) [box, right of=channel2, xshift=1.5cm,  font=\fontsize{12}{12}\selectfont] {Decoder2\\($g_{\bm{\phi_2}}$)};
\node (xhat2) [txt, right of=decoder2,font=\fontsize{14}{0}\selectfont] {$\bm{\hat{x}_2}$};
\node (reflayer) [fitted, fit=(channel2) (decoder2)] {};
\node at (reflayer.south) [below, inner sep=1mm] {Refinement Layer};


\draw [arrow] (x) -- (encoder);
\draw [arrow] (encoder.east) 
-- node[above,font=\fontsize{14}{0}\selectfont] {$\bm{z}$} ++(1, 0) |- node[above right,font=\fontsize{14}{0}\selectfont] {$\bm{z_1}$}  (channel1.west);
\draw [arrow] (channel1) -- node[above,font=\fontsize{14}{0}\selectfont] {$\hat{\bm{z}}_1$} (decoder1);
\draw [arrow] (decoder1) -- (xhat1);
\draw [arrow] (encoder.east) 
-- ++(1, 0) |- node[below right,font=\fontsize{14}{0}\selectfont] {$\bm{z_2}$}  (channel2.west);
\draw [arrow] (channel2.350) -- node[above left,font=\fontsize{14}{0}\selectfont] {$\hat{\bm{z}}_2$} (decoder2.187);
\draw [arrow] (channel1.east) -- ++(0.75, 0) |- (decoder2.165);
\draw [arrow] (decoder2) -- (xhat2);


\end{tikzpicture}

}%
	\end{center}
  \caption{DeepJSCC-$l$ architecture for progressive wireless image transmission with two layers, performing successive refinement. An input image is encoded into layers $\bm z_1$ and $\bm z_2$, each of them transmitted over different realizations of the noisy channel.
  }
  \label{fig:modeldirectpass}
\end{figure}
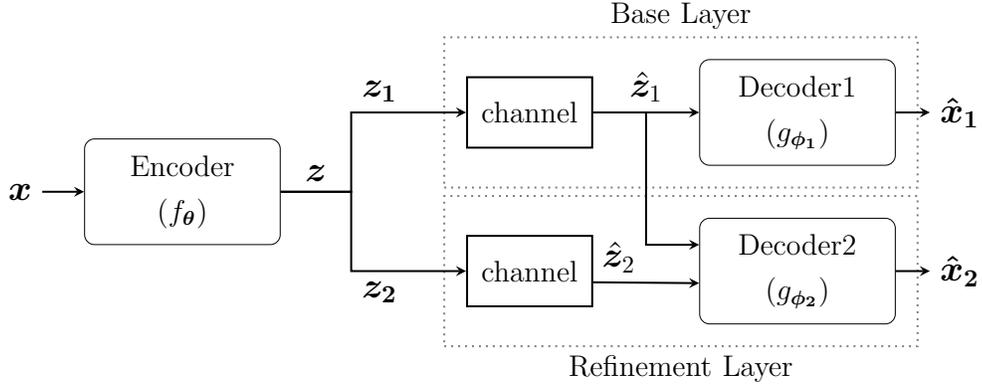

We start with the successive refinement problem, in which the decoder receives the outputs of the first $i$ channels for some $1 \leq i \leq L$.
We refer to the symbols transmitted over the first channel as the \emph{base layer}, and the following channels as the \emph{refinement layers}.

The first solution is based on an architecture consisting of a single encoder NN and $L$ independent decoder NNs, as illustrated in Fig.~\ref{fig:modeldirectpass} for $L=2$. 
The whole system is modeled as an autoencoder and all the layers are trained jointly, with the loss function defined as:

\begin{equation}
  \mathcal{L} = \frac{1}{L}\frac{1}{N}\sum_{j=1}^L\sum_{i=1}^N d(\bm x^i, \bm{\hat x}_{j}^{i}),
  \label{eq:lossdirect}
\end{equation}
where $d(\bm x^i, \bm{\hat x}_{j}^{i})$ is the MSE distortion between the original image $\bm x^i$ and its reconstruction at decoder $j$, $\bm{\hat{x}}_j^i$, for the $i$-th sample of the training dataset, and $N$ is the number of training samples. Note that the loss function in (\ref{eq:lossdirect}) puts equal weights on the distortions of all the $L$ decoders. Although a more general loss function could be formulated with different weights per distortion achieved by different decoders, experimental results showed that this has marginal impact on the performance. For more details, please see Appendix \ref{sec:weights}.

\begin{figure}
    \centering
    \includegraphics{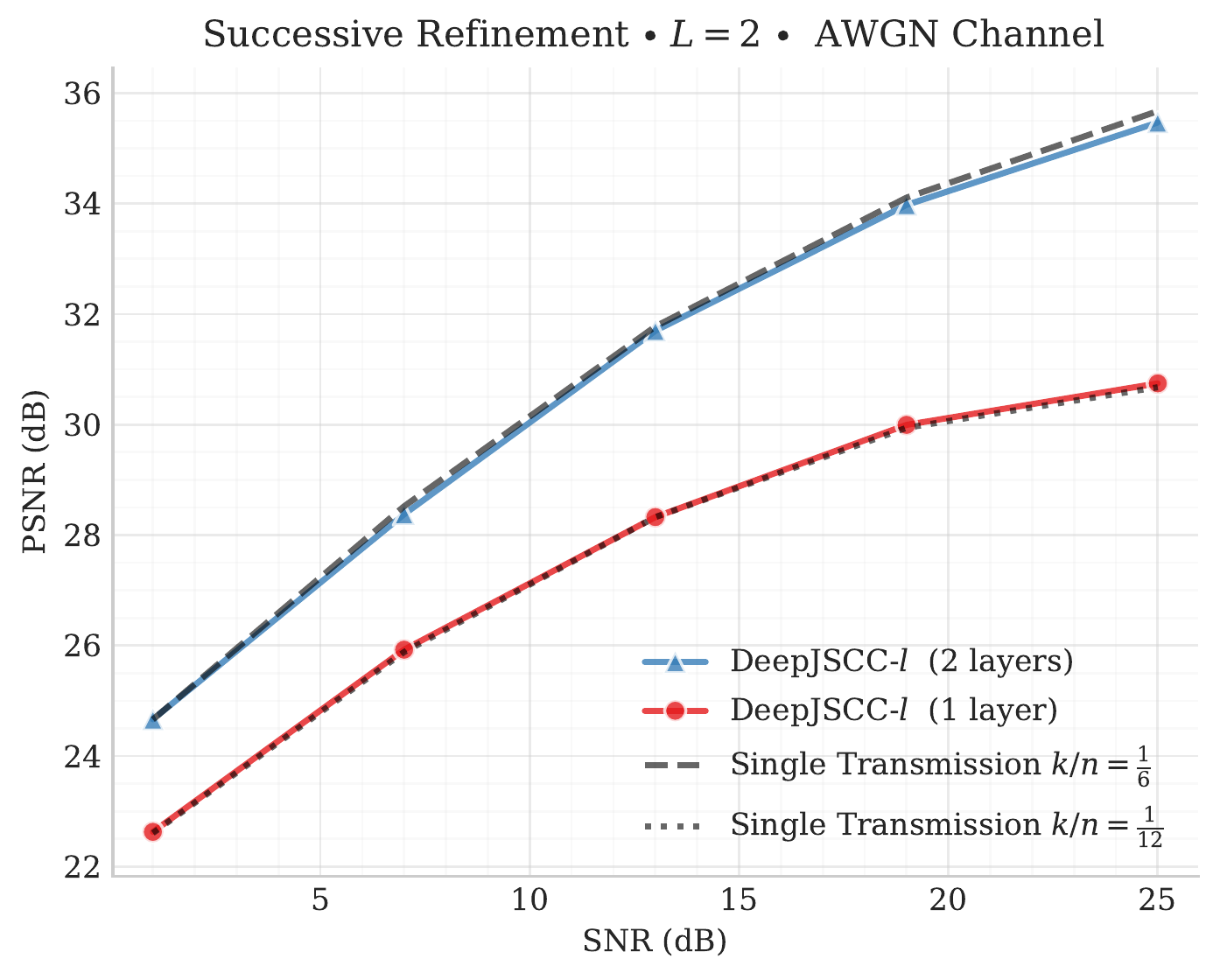}
    \caption{DeepJSCC-$l$ performance for successive refinement with $L=2$ layers over a wide range of SNRs, for $k_1/n = k_2/n = 1/12$. Colored curves show the performance of reconstructions using both subsets of channel outputs ($\bm{\hat{x}}_1$ and $\bm{\hat{x}}_2$). Black dashed lines plot the performance of the single transmission model with equivalent bandwidth. Our results show that the loss due to layering is negligible.}
    \label{fig:dpass-hull}
\end{figure}

\subsection{Two-layer Model}

Our first set of results focus on the $L=2$ layers scenario, which requires the training of only one encoder and two decoders.
We consider $k_1/n = k_2/n = 1/12$, and the AWGN channel. In Fig.~\ref{fig:dpass-hull}, we present the results for different channel SNRs, where each point in the figure is achieved by training a distinct encoder-decoder pair.
As a comparison baseline, we also present the performance achieved by the DeepJSCC scheme with a single layer using the same bandwidth as ${\bm{\hat Z}}_1$ and ${\bm{\hat Z}}_2$ ($k/n = 1/12$ and $k/n = 1/6$), respectively.

For all the channel conditions, the average $\mathrm{PSNR}_2$ is consistently higher than $\mathrm{PSNR}_1$ by 2 to 3 dB, showing the contribution of the refinement layer.
The results also demonstrate that DeepJSCC-$l$ 
can learn to transmit a sequential representation of the input images, while maintaining the performance close to the baseline curves. 
The fact that the performance loss compared to the baseline is negligible implies that DeepJSCC-$l$ is able to find a nearly successively refinable representation over Gaussian channels; that is, the flexibility of allowing a decoder to reconstruct the imaged based only on the base layer, or both the layers comes at almost no cost in performance.

\subsection{Adaptability to Varying Channels}

\begin{figure*}
	\begin{center}
 		\subfloat[]{\includegraphics[width=0.5\textwidth]{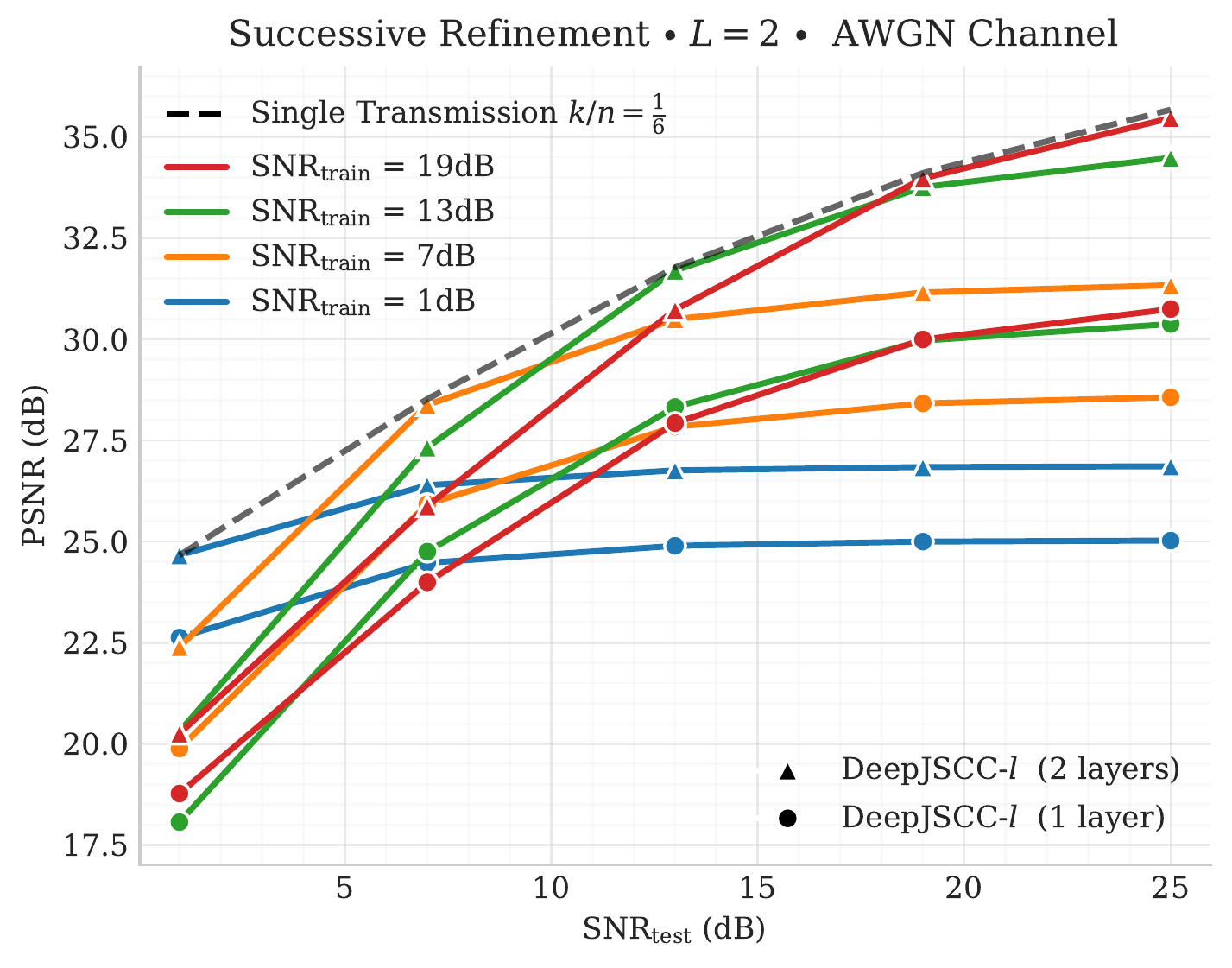} \label{fig:dpass-awgn-05}}
		\subfloat[]{\includegraphics[width=0.5\textwidth]{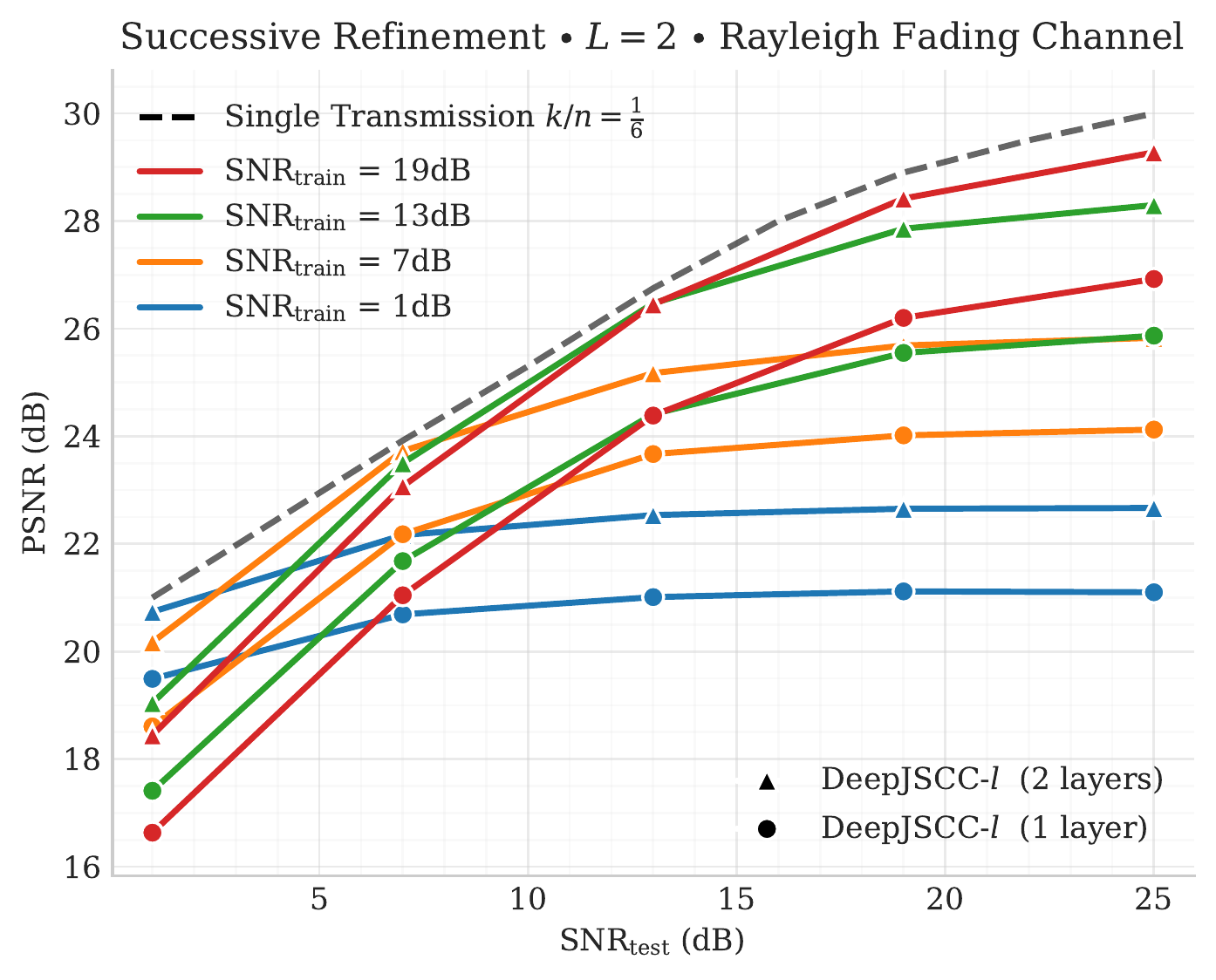}\label{fig:dpass-fading}}
	\end{center}
  \caption{DeepJSCC-$l$ performance on successive refinement when there is disparity between training and test channel conditions, typical from multi-user communication. Each color represents the performance over a range of SNR for a DeepJSCC-$l$ model trained for a specific SNR; triangle markers correspond to receivers using $k_1/n = 1/12$ bandwidth ratio (base layer), while circle markers correspond to receivers using $k_1/n + k_2/n = 1/6$ bandwidth ratio (base+refinement layers). Two channel models are considered: (a) AWGN channel and (b) slow Rayleigh fading channel.}
\end{figure*}

A common issue in real systems is the mismatch of conditions between design and deployment stages. Often a system is designed having a specific target communication channel condition, but when deployed the channel conditions may have changed. Also, most practical systems rely on imperfect channel estimation and feedback, which results in a mismatch between the channel state assumed at the transmitter (and used for picking the rate for compression and channel coding) and the real channel condition.
This can be a serious issue for digital systems, as significant mismatch can lead to a total loss of information at the receiver, known as the \emph{cliff effect}.

To simulate such a scenario, we consider the performance of DeepJSCC-$l$ trained on a specific target SNR, but evaluated at a range of different channel conditions. Fig. \ref{fig:dpass-awgn-05} shows the performance results for both the base layer transmission and the base+refinement layers, where each color represents the performance of a model trained for a specific SNR, with the curve with circle markers corresponding to the performance of the decoder receiving only the base layer, while the one with triangle markers the decoder receiving both layers. As reference, we also plot with the black dashed line the best performance at different SNRs of a single-layer DeepJSCC.

The results show that the performance of DeepJSCC-$l$ deteriorates gradually (yet not abruptly) for both reconstructions when the test SNR is lower than the trained SNR, showing that it is robust against SNR mismatch. Similarly, unlike in digital systems, the performance of DeepJSCC-$l$ for both reconstructions improves gradually with the channel SNR. This shows that DeepJSCC-$l$ does not suffer from the cliff effect but instead presents \emph{graceful degradation}. Note that this is a behavior typical for analog systems and was already observed in the single layer case in \cite{bourtsoulatze:TCCN:2019}.
We also observe that the performance gap between the layers tend to remain constant when the test SNR is higher than the trained SNR, but the gap reduces as the test SNR falls below the training SNR. This is because the benefit from the refinement layer degrades as the test SNR decreases since the reconstructed base layer becomes significantly different from what the encoder expects based on the training data.

We also train DeepJSCC-$l$ over a slow Rayleigh fading channel, when the channel realization remains constant for the duration of the transmission of each layer, but takes an independent value for the transmission of each image. This scenario can also represent a multi-user multicasting scenario, in which a different ``virtual'' receiver corresponds to each realization of the channel.

Fig.~\ref{fig:dpass-fading} shows the results for the same model architecture as in Fig.~\ref{fig:dpass-awgn-05}, where the x-axis denotes the average SNR in the test phase. We see that, although the PSNR values are lower than those in the AWGN case due to channel uncertainty, the properties of graceful degradation and limited loss with respect to the single-layer baseline are preserved.

We highlight here that DeepJSCC-$l$ does not exploit explicit pilot signals or channel estimation, yet it is able to adapt to the channel uncertainty.
All the models presented in this paper exhibit similar behavior of graceful degradation and capacity to learn over fading channels. In the remainder of this paper, we only present the highest PSNR obtained for each channel SNR value, and only consider transmissions over an AWGN channel.

\begin{remark}
We remark here that, due to the analog nature of DeepJSCC-$l$, the reconstruction at the receiver based on the first $l$ layers is not fixed, and depends on the realization of the random channel. Therefore, unlike in digital systems, the exact reconstruction at the decoder cannot be known by the encoder in advance; and hence, the second layer cannot simply transmit the residual information. It is remarkable that DeepJSCC-$l$ can learn to refine the previous reconstructions despite this uncertainty, even in the case of a fading channel.
\end{remark}

\subsection{Multiple Layers}

\begin{figure*}
	\begin{center}
    \subfloat[]{\includegraphics[width=0.5\textwidth]{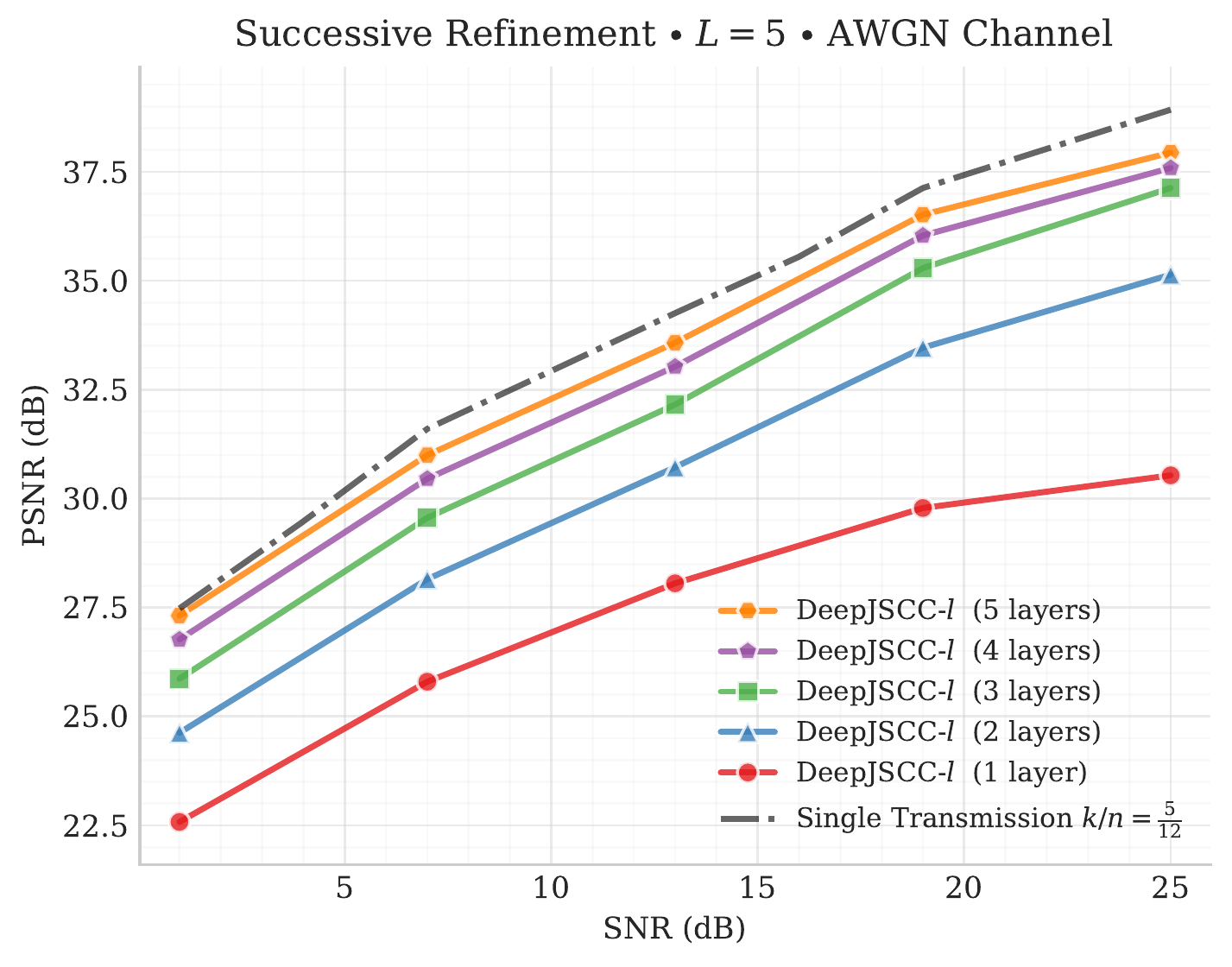}\label{fig:dpass-mult}}
    \subfloat[]{\includegraphics[width=0.5\textwidth]{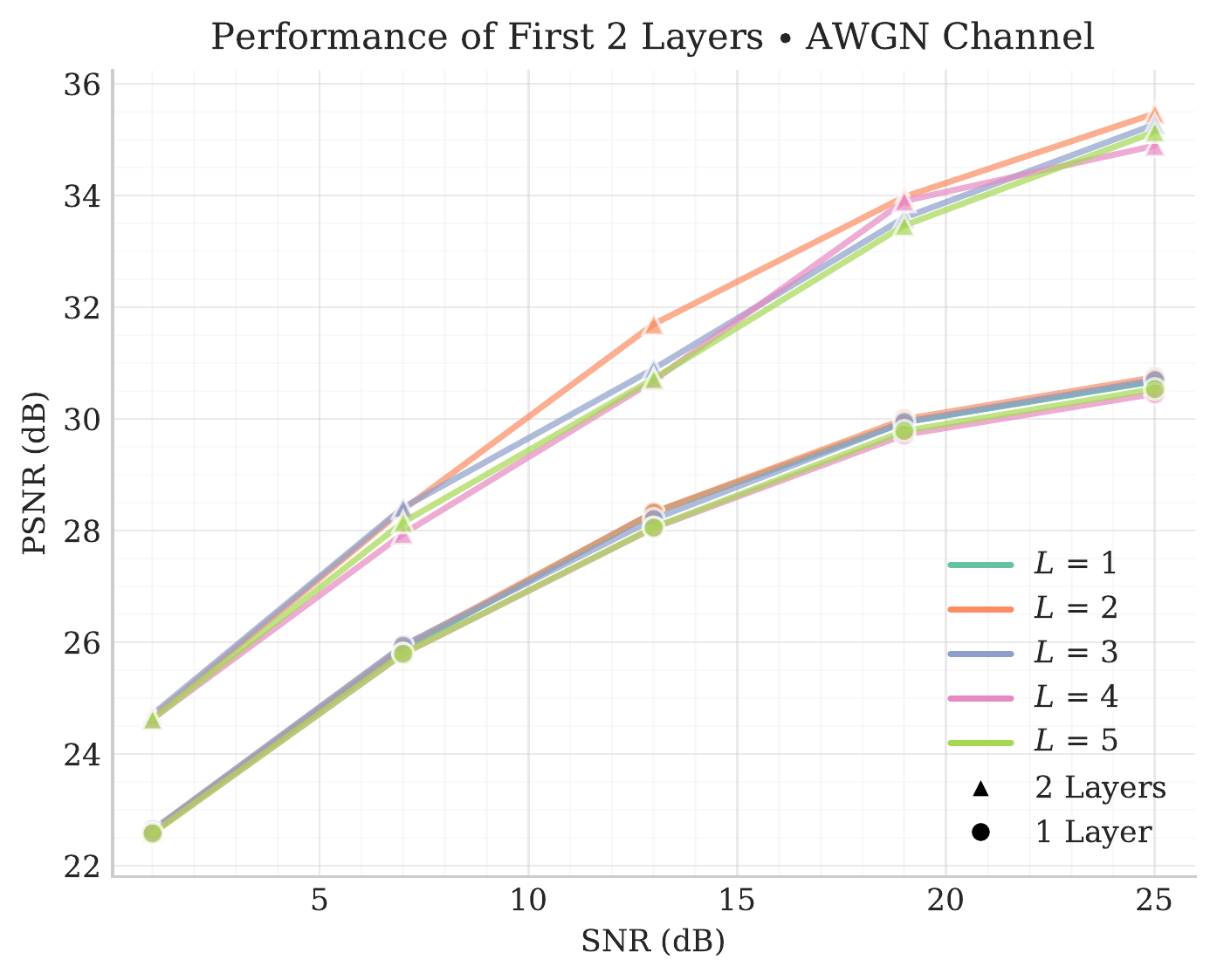}\label{fig:dpass-laycomp}}
	\end{center}
  \caption{(a) Performance of DeepJSCC-$l$ using $L=5$ layers over different SNRs. Note that the increase in performance with each refinement layer gradually decreases. (b) Performance of the two first layers ($\bm{\hat{x}}_1$ and $\bm{\hat{x}}_2$) for DeepJSCC-$l$ trained with different values of $L$. Note that despite the increase in the number of layers, the performances of the first two layers remain relatively stable. In both plots, $k_i/n = 1/12,~ \forall i \in \{1 \dots 5\}$.}
\end{figure*}

Next, we extend the model to more than two layers. Fig.~\ref{fig:dpass-mult} shows the results for $L=5$ layers, each transmitted with a bandwidth ratio equal to $1/12$.
The results show that the addition of new layers increases the overall quality of the transmitted image at every step; although the amount of improvement is diminishing, as the model is able to transmit the main image features with the lower layers, leaving only marginal contributions to the additional layers.

We also notice that the introduction of additional layers in the training model has very low impact on the performance of the first layers, compared to models with smaller values of $L$. This can be seen in Fig.~\ref{fig:dpass-laycomp}, which compares the performance of the first and second layers for models trained with $L \in \{2,3,4,5\}$, showing that the loss due to the addition of new layers is negligible.
This is rather surprising, given that the code of the first layer is shared by all the layers and is optimized to be maximally useful in combination with a number of refinement layers. The results, therefore, suggest that there is almost performance independence between layers, justifying the use of as many layers as desired, as long as there are available resources.

\subsection{Comparison with Digital Transmission}
\label{sec:digital}

\begin{figure}
	\begin{center}
 		\subfloat[]{\includegraphics[width=0.49\textwidth]{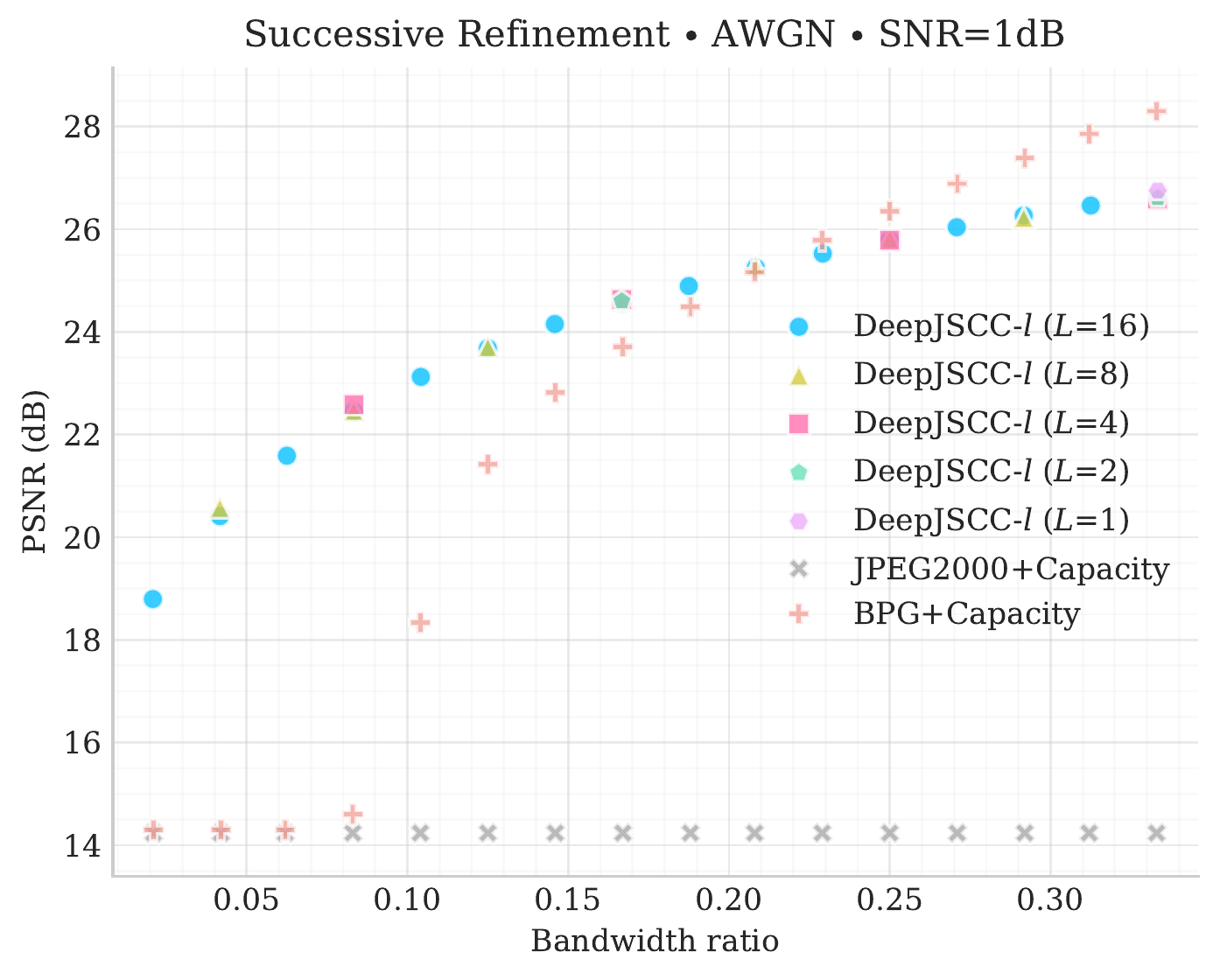}\label{fig:range1}} 
		\subfloat[]{\includegraphics[width=0.49\textwidth]{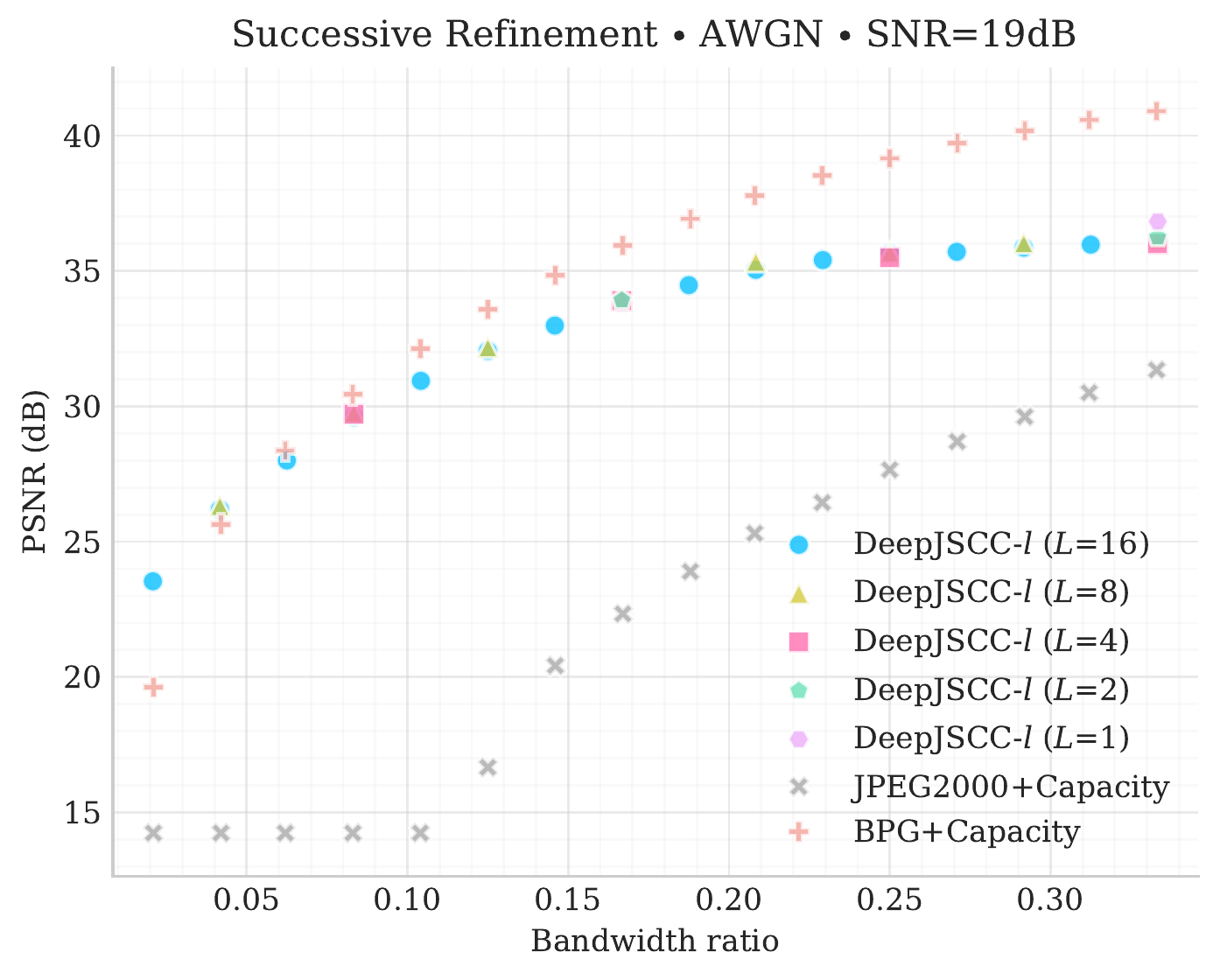}\label{fig:range19}}
	\end{center}
  \caption{PSNR vs.\ bandwidth ratio comparison for $L = 1, 2, 4, 8$ and $16$ layers at (a) SNR = 1dB and (b) SNR = 19dB. DeepJSCC-$l$ presents superior performance for the first layers when compared to a separation-based scheme using JPEG2000 (with 16 layers) or BPG for compression, and an ideal capacity-achieving code.}\label{fig:range}
\end{figure}

Finally, we consider an experiment in which a fixed bandwidth $k$ is divided into $L$ layers of equal size. Fig.~\ref{fig:range} shows the results of five different cases corresponding to $L = 1, 2, 4, 8, 16$, and total bandwidth ratio $k/n = 1/3$ for $\mathrm{SNR}=1$dB (Fig.~\ref{fig:range1}) and $\mathrm{SNR}=19$dB (Fig.~\ref{fig:range19}). The performance of all the reconstructions for each model are shown. We observe that there is almost no loss in performance by dividing the transmission into many layers, as many as $L=16$, while this provides additional flexibility, i.e., a receiver may stop receiving after having received a certain number of layers if it has reached a certain target quality, and may use the bandwidth and processing power for other tasks.

For comparison, we also consider digital transmission employing separate source and channel codes.
In particular, we consider both JPEG2000\footnote{\url{https://jpeg.org/jpeg2000/index.html}} and BPG\footnote{\url{https://bellard.org/bpg/}} as the source encoder, followed by a capacity-achieving channel code. JPEG2000 is chosen as it can to generate layered representations at different bit rates; the choice of BPG is motivated by its high performance in image compression. The capacity-achieving channel code is an ideal formulation assuming that bits can be transmitted without errors at the channel capacity \cite{Shannon:IT:56}. Although near capacity-achieving channel codes exists for the AWGN channel, what we are assuming here is not feasible in practice for the blocklengths considered here.
Thus, this scheme would serve as an upper bound on the performance of any separation based digital scheme employing JPEG2000 or BPG for compression.

The digital scheme works as follows. For a given bandwidth ratio $k_i/n$, source dimension $n$, and the channel capacity at each SNR, a bit budget $b_i$ is determined as the maximum number of bits that can be transmitted over $k_i$ channel uses. When using JPEG2000 as source code, we compress images into $L$ layers, each using at most $b_i$ bits. For BPG, as the official encoder does not allow layered compression, we produce independent compressions with the best possible quality for each $b_i$ target.
For fair comparison we discard the bits dedicated to header, so only the compressed pixels are transmitted.

The results show that DeepJSCC-$l$ can achieve performance superior than or comparable with JPEG2000 and BPG codecs, particularly for the lower layers. The improvement is particularly noticeable in Fig.~\ref{fig:range1} where the low SNR ($1$dB) and the constrained bandwidth ratio limits the channel capacity so much that the JPEG2000 and BPG codecs are unable to compress images to the level supported by the channel, resulting in the flat curve regions displayed in the graph.

\begin{figure}
	\begin{center}
 	\includegraphics[width=0.45\textwidth]{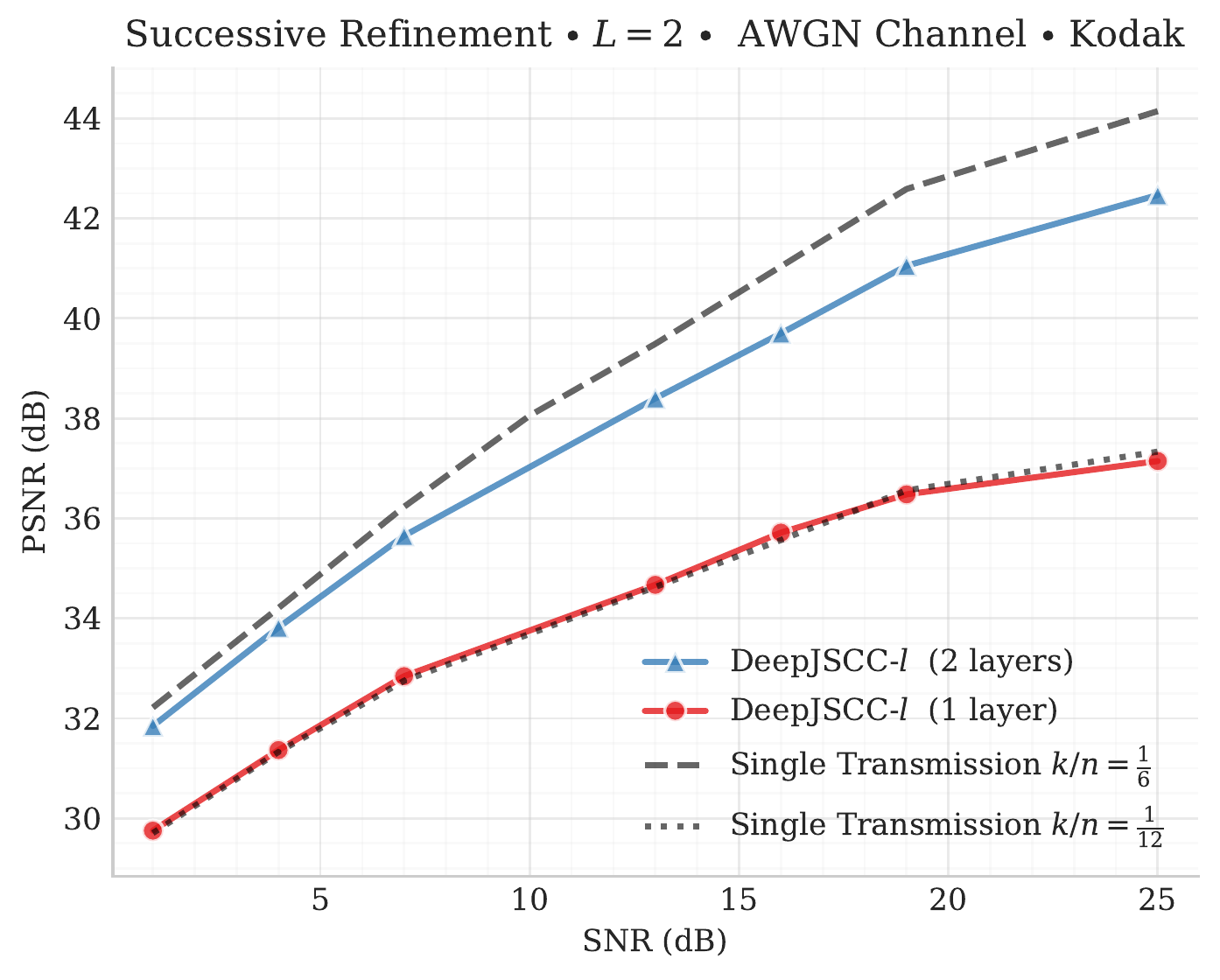} \label{fig:multdec-kodak}
	\end{center}
		\caption{Performance results with encoder/decoder architecture introduced in \cite{kurka:IZS:2020}. Model was trained on ImageNet dataset and evaluated on Kodak dataset. 
		}\label{fig:kodak}
\end{figure}

\begin{figure}
	\begin{center}
 		\subfloat[Original image]{\includegraphics[width=0.33\textwidth]{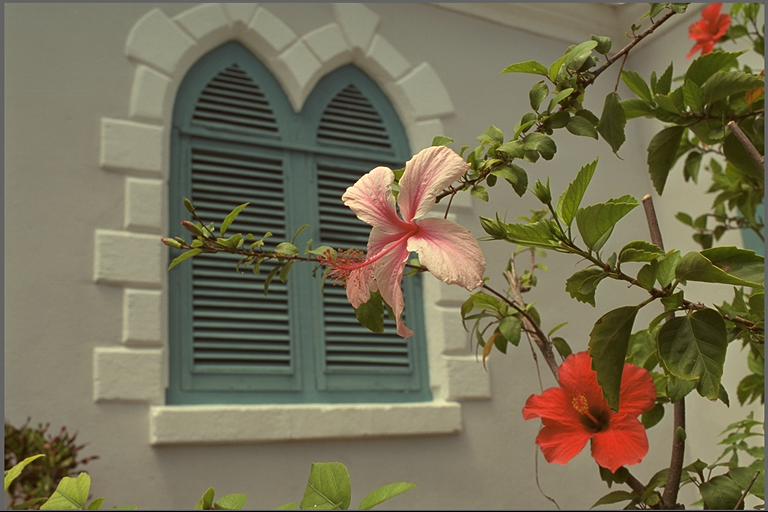}\label{fig:visual_original}} 
		\subfloat[Base reconstruction, $\bm{\hat x}_1$\protect\\PSNR: 31.90 /
		MS-SSIM: 0.9749
		]{\includegraphics[width=0.33\textwidth]{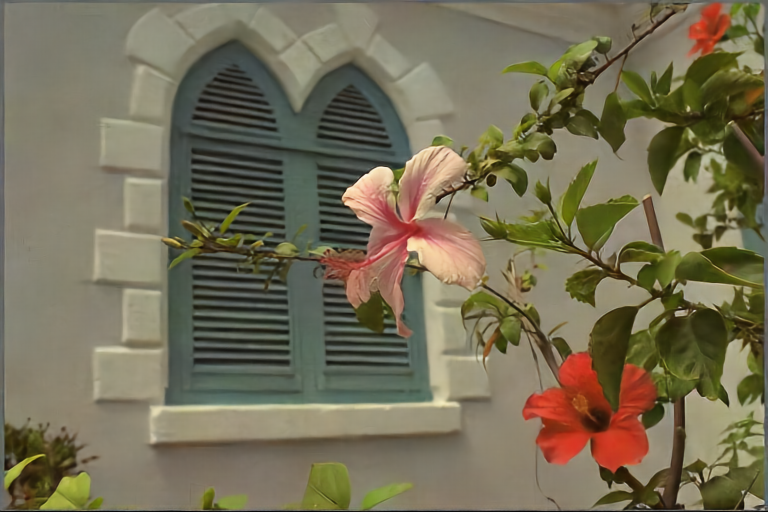}\label{fig:visual_layer1}}
		\subfloat[Refined reconstruction, $\bm{\hat x}_2$\protect\\PSNR: 34.45 /
		MS-SSIM: 0.9861]{\includegraphics[width=0.33\textwidth]{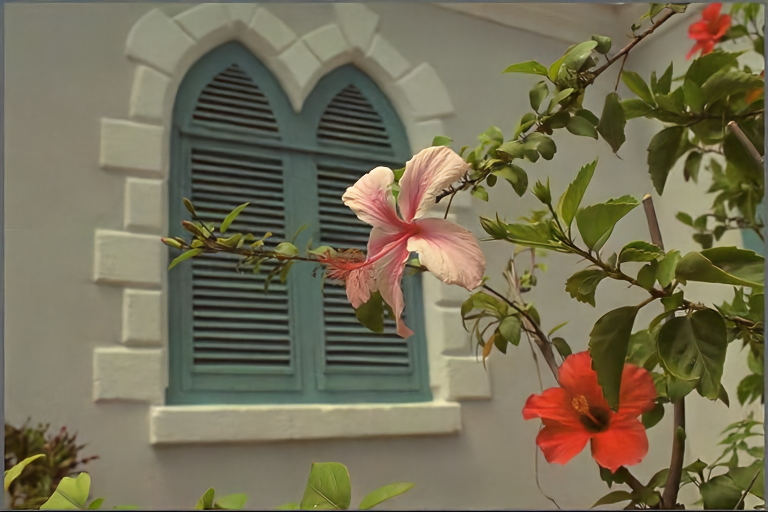}\label{fig:visual_layer2}}
		\\
		\subfloat[Highlight $\bm{\hat x}_1$]{\includegraphics[width=0.4\textwidth]{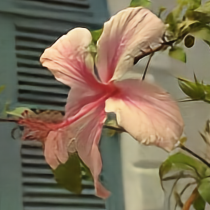}\label{fig:visual_layer1_high}}
				\subfloat[Highlight $\bm{\hat x}_2$]{\includegraphics[width=0.4\textwidth]{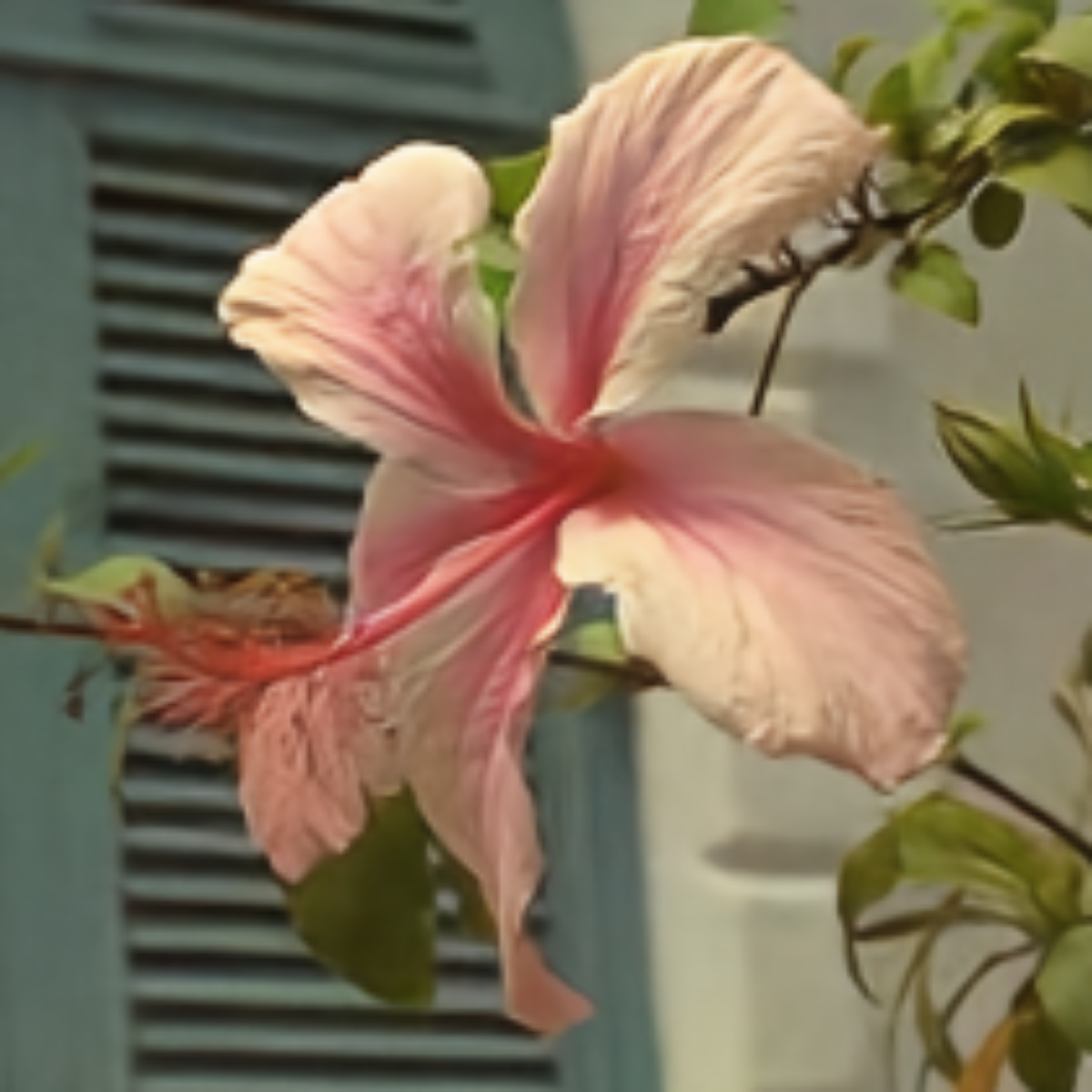}\label{fig:visual_layer2_high}}
	\end{center}
  \caption{Examples of reconstructions of successive refinement model. Note how flower details are enhanced between Figures (d) and (e).}\label{fig:visual-sucref}
\end{figure}

We have also experimented extending DeepJSCC-$l$ with different layers and hyperparameters, and training the network on larger datasets. 
In particular, we experimented using the network hyperparameters in \cite{kurka:IZS:2020} for the encoder and decoder, which uses bigger convolutional filters and apply generalized normalization transformations \cite{balle2015density} inspired by the sate-of-the-art for neural image compression models in \cite{BalleICLR2018,Balle:NIPS2018,minnen2020channelwise}. 
The models were trained on the ImageNet dataset, consisting of more and larger resolution images than CIFAR-10. The model is then evaluated on the Kodak dataset and the PSNR curves compared to the single-layer model for $L=2$ is provided in Fig.~\ref{fig:kodak}, while examples of reconstructions are presented in Fig.~\ref{fig:visual-sucref}.
Although these more complex network architectures produce remarkable performance (see \cite{kurka:IZS:2020} for a detailed comparison with many codecs), they also require significantly larger training time, particularly for multi-layer compression tasks, which is why we have limited our numerical results for more layers to the simpler architecture of~\cite{bourtsoulatze:TCCN:2019}.

\subsection{Alternative Architectures}

The model architecture for DeepJSCC-$l$ introduced in Fig.~\ref{fig:modeldirectpass} does not represent the only viable solution for the successive refinement problem.
Here, we discuss alternative DeepJSCC-$l$ architectures with different trade-offs. We note that the trade-off is between the space and time complexity, and not necessarily the performance, as all the methods we present below achieve comparable performance to the one presented above.

\begin{figure*}
	\begin{center}
		\subfloat[]{\includegraphics[width=0.45\textwidth]{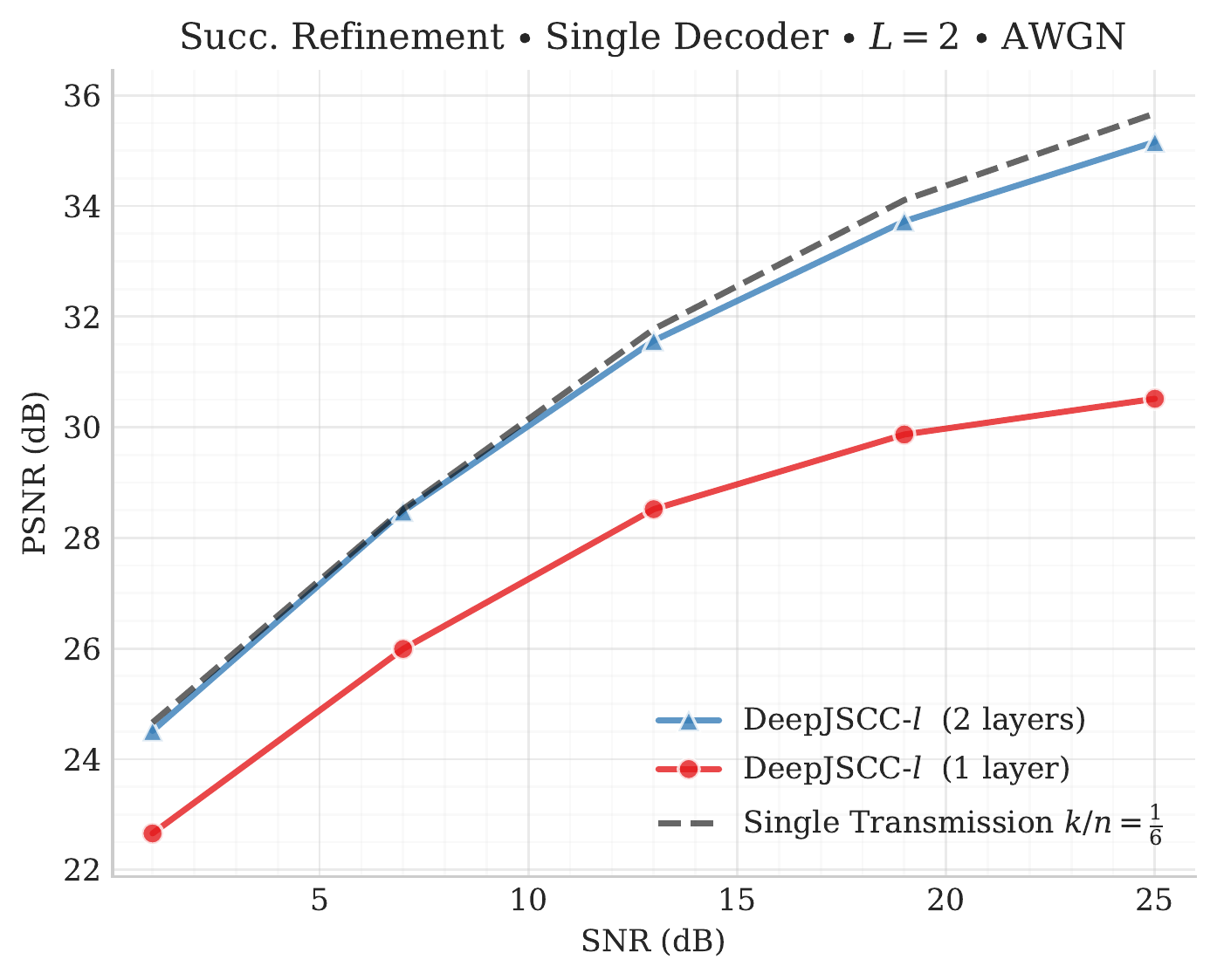}\label{fig:single2}}
 		\subfloat[]{\includegraphics[width=0.45\textwidth]{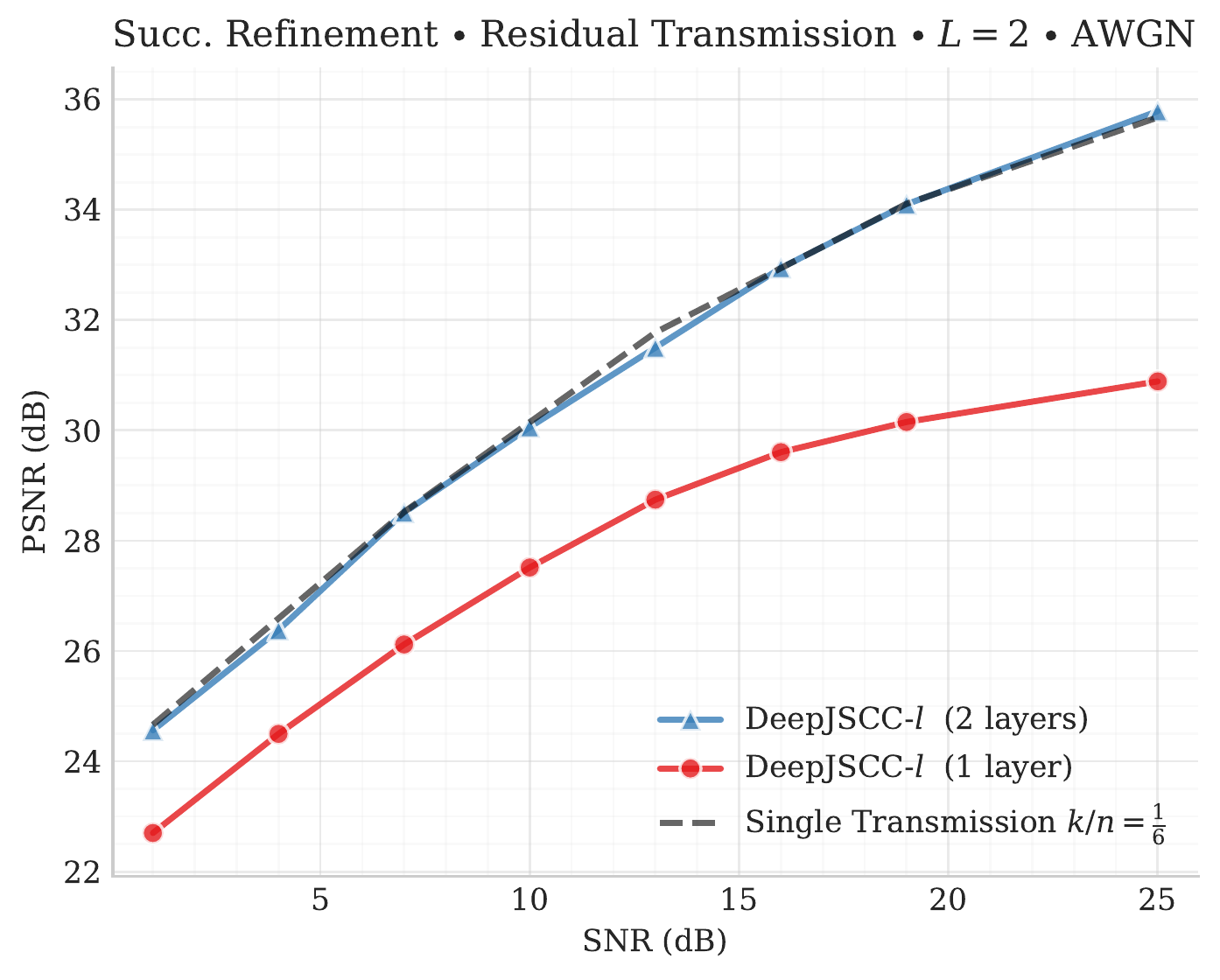} \label{fig:resnet-awgn}}\\
	\end{center}
		\caption{Performance of alternative successive refinement DeepJSCC-$l$ architectures on CIFAR-10 test images, transmitted over an AWGN channel with $k_1/n = k_2/n = 1/12$. (a) Single decoder scheme, (b) Residual transmission,  $m=10$. Both architectures have performance equivalent to the first scheme proposed (multiple decoders), but presenting different space and time complexities.
		}
\end{figure*}

\subsubsection{Single Decoder}
\label{sec:muldec}

\begin{figure}
	\begin{center}
	    \includegraphics[width=0.85\linewidth]{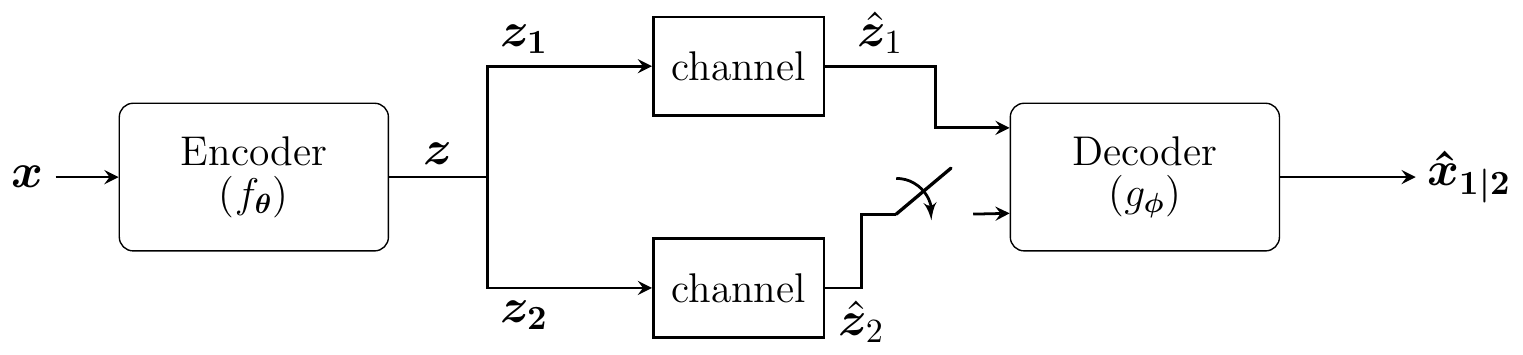}
	\end{center}
\caption{Single decoder scheme with two layers. A single decoder is trained with different input sizes, being able to reconstruct the image with as many layers as it is provided with.}
\label{fig:modelsingdec}
\end{figure}

A downside of the model used previously (Fig.~\ref{fig:modeldirectpass}) is the fact that a separate decoder needs to be trained for each layer. Here we try an alternative model that uses a single encoder and a single decoder architecture for all the layers, as illustrated in Fig.~\ref{fig:modelsingdec}.
In order to retrieve information from partial subsets, the decoder has to be trained for different input sizes. We achieve that by exposing a single decoder to different code lengths, and averaging its performance over all possible subsets of layers.
In practical terms, that means creating a CNN model with fixed channel bandwidth $k = \sum_{i=1}^L k_i$, but randomly masking consecutive regions of size $k_i$ from the end of the received message $\bm{\hat z}$ with zeros. In this way, the network should learn to specialize different regions of the code, using the initial parts to encode the main image content and the extra (occasionally erased) parts for additional layers.
This can also be considered as ``structured dropout'', where the dropout during training allows training a decoder that can adapt to the available bandwidth.

Note that during training, the length of the transmitted code (i.e., the number of layers) is defined randomly at every batch. This is essential so that the encoder and decoder can preserve the performance of all the layers. An alternative approach that train subsets of layers sequentially until convergence with sizes 1 to $L$ showed to be detrimental to the performance of the first layers. This happened because the training of higher order layers modified the parameters of previous layers.

The results presented in Fig.~\ref{fig:single2} for $L=2$ layers show that the performance of DeepJSCC-$l$ with a single decoder is close to the single transmission bound. The achieved values are as good as in the multiple decoder architecture (Fig.~\ref{fig:dpass-hull}).
This model is particularly appealing as it represents a considerable reduction both in memory and processing as the model size remains the same regardless of the number of layers. However, while the multiple decoder scheme learns separate decoders for all the layers in parallel, the single decoder strategy has to be presented with different codelengths, increasing the training time.

\subsubsection{Residual Transmission}
\label{sec:resnet}

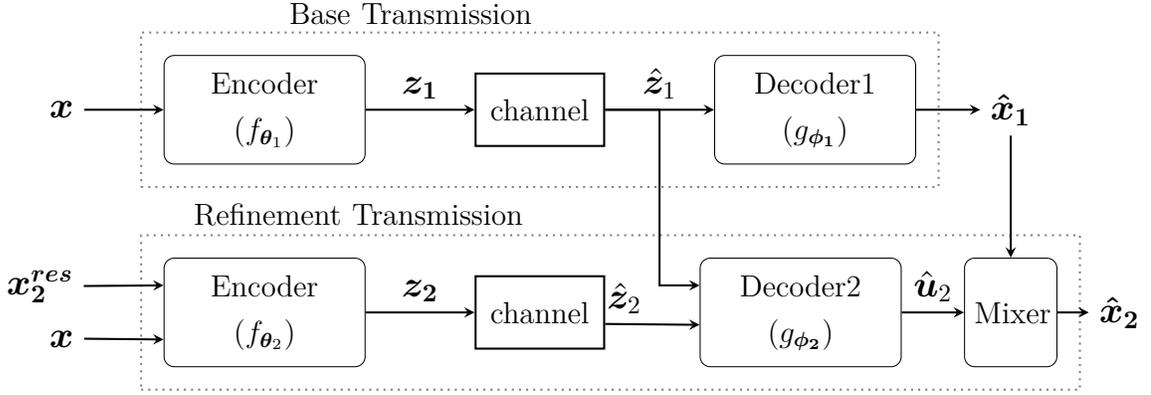
\begin{figure}
	\begin{center}
        \resizebox {\linewidth} {!}%
        {

 \tikzstyle{txt} = [text centered]
 \tikzstyle{box} = [rectangle, rounded corners, minimum width=2.5cm, minimum height=1.5cm, text width=2.5cm, text centered, draw=black]
 \tikzstyle{smallbox} = [rectangle, rounded corners, minimum width=1cm, minimum height=1.5cm, text width=1cm, text centered, draw=black]
 \tikzstyle{bbox} = [rectangle, thick, minimum width=1.5cm, minimum height=1cm, text width=1.5cm, text centered, draw=black]
 \tikzstyle{arrow} = [thick,->,>=stealth]
 \tikzstyle{fitted} = [draw=gray, thick, dotted, inner sep=0.75em]
 \tikzstyle{sum} = [draw, circle]


\begin{tikzpicture}[node distance=2.3cm]

\node (x1) [txt, font=\fontsize{14}{0}\selectfont] {$\bm{x}$};
\node (encoder1) [box, right of=x1, xshift=0.5cm, font=\fontsize{12}{12}\selectfont] {Encoder \\ ($f_{\bm{\theta}_1}$)};
\node (channel1) [bbox, right of=encoder1, xshift=1cm, font=\fontsize{12}{12}\selectfont, xshift=0.5cm] {channel};
\node (decoder1) [box, right of=channel1, xshift=1.5cm,  font=\fontsize{12}{12}\selectfont] {Decoder1\\($g_{\bm{\phi_1}}$)};
\node (xhat1) [txt, right of=decoder1, xshift=0.4cm, font=\fontsize{14}{0}\selectfont] {$\bm{\hat{x}_1}$};
\node (baselayer) [fitted, fit=(encoder1) (decoder1)] {};
\node at (baselayer.north) [above left, inner sep=1mm] {Base Transmission};


\node (encoder2) [box, below of=encoder1, yshift=-0.5cm, font=\fontsize{12}{12}\selectfont] {Encoder \\ ($f_{\bm{\theta}_2}$)};
\node (x2) [txt, left of=encoder2, xshift=-0.8cm, yshift=0.36cm, font=\fontsize{14}{0}\selectfont] {$\bm{x^{res}_2}$};
\node (x22) [txt, left of=encoder2, xshift=-0.5cm, yshift=-0.36cm, font=\fontsize{14}{0}\selectfont] {$\bm{x}$};
\node (channel2) [bbox, right of=encoder2, xshift=1cm, font=\fontsize{12}{12}\selectfont, xshift=0.5cm] {channel};
\node (decoder2) [box, right of=channel2, xshift=1.3cm, font=\fontsize{12}{12}\selectfont] {Decoder2\\($g_{\bm{\phi_2}}$)};
\node (adder) [smallbox, right of=decoder2, xshift=0.6cm] {Mixer};
\node (xhat2) [txt, right of=adder, xshift=-0.8cm, font=\fontsize{14}{0}\selectfont] {$\bm{\hat{x}_2}$};
\node (reflayer) [fitted, fit=(encoder2) (adder)] {};
\node at (reflayer) [above left, inner sep=12mm] {Refinement Transmission};


\draw [arrow] (x1) -- (encoder1);
\draw [arrow] (encoder1) -- node[above,font=\fontsize{14}{0}\selectfont] {$\bm{z_1}$} (channel1);
\draw [arrow] (channel1) -- node[above,font=\fontsize{14}{0}\selectfont] {$\hat{\bm{z}}_1$} (decoder1);
\draw [arrow] (decoder1) -- (xhat1);
\draw [arrow] (x2) -- (encoder2.165);
\draw [arrow] (x22) -- (encoder2.195);
\draw [arrow] (channel2.350) -- node[above left,font=\fontsize{14}{0}\selectfont] {$\hat{\bm{z}}_2$} (decoder2.187);
\draw [arrow] (channel1.east) -- ++(0.75, 0) |- (decoder2.165);

\draw [arrow] (decoder2) -- node[above,font=\fontsize{14}{0}\selectfont] {$\hat{\bm{u}}_2$} (adder);
\draw [arrow] (xhat1) --  (adder);
\draw [arrow] (adder) -- (xhat2);
\draw [arrow] (encoder2) -- node[above,font=\fontsize{14}{0}\selectfont] {$\bm{z_2}$}  (channel2.west);

\end{tikzpicture}

}%
	\end{center}
    \caption{Residual transmission scheme with two layers. At each layer, the residual of the previous transmissions is estimated and transmitted. Additional layers can be introduced without the need to retrain existing layers.}\label{fig:modelresnet}
\end{figure}

Another alternative architecture we propose is based on residual transmission. Here, as illustrated in Fig.~\ref{fig:modelresnet}, each transmission is performed by an independent encoder/decoder pair acting sequentially. Instead of jointly optimizing all the parameters of all the layers simultaneously, we use a greedy approach in which an encoder/decoder pair is trained until convergence and their weights are fixed (frozen) so new pairs can be trained on top of it.

The first encoder/decoder pair (the base layer) behaves exactly as in the single transmission scheme, transmitting the original image $
\bm x$, compressed at rate $k_{1}/n$, and retrieved as $\bm{\hat x}_{1}$. 
Then, in each subsequent layer $j$, the encoder uses as input the original image being transmitted, $\bm x$, and an estimate of the residual error between the original image and its estimate of the receiver based on the previous $j-1$ layers, $\hat {\bm x}_{j-1} $,

\begin{equation*}
  {\bm x}_j^{\textit{res}} \triangleq {\bm x} - {\bm{\hat x}'_{j-1}}.
  \label{eq:input-resnet}
\end{equation*}
Here, since the transmitter does not know the reconstructed image at the receiver, $\bm{\hat x'}_{j-1}$ is an estimate of $\bm{\hat x}_{j-1}$ based on the statistics of the dataset and the channel.

We assume the transmitter has a local copy of the decoder parameters at previous layers. So, in order to generate $\bm{\hat x'}_{j-1}$, the transmitter simulates locally independent realizations of the channel and the decoder models, obtaining

\begin{equation*}
  \label{eq:resnet-estimated-input}
  {\bm{\hat x}'_{j-1}} = \frac{1}{m}\sum_{i=1}^m {\bm{\tilde x}^i_{j-1}},
\end{equation*}
where, with abuse of notation, $\bm{\tilde x}^i_{j-1}$ is the $i$-th realization of the simulation of the transmitter's image reconstruction, and $m$ is the total number of independent channel realizations used to estimate the receiver's output.
Note that this estimation at the transmitter side is necessary because we assume no feedback channel between the receiver and transmitter. In the presence of feedback, the receiver's reconstruction could be sent back to the transmitter~\cite{kurka_deepjscc-f_2020}.

In the residual transmission scheme, each layer $i > 1$ encodes and decodes an estimated residual image, containing the missing information not transmitted yet, that can be combined with the reconstruction at $i-1$, producing the refinement. The combination of the previous reconstruction and refinement is done by the decoder network. At layer $i$, the decoder $i$ receives as input the concatenation of all the channel outputs received so far to reconstruct a residual estimate $\bm{\hat u}_i$. Later, $\bm{\hat u}_i$ is combined with the reconstruction at the previous layer $\bm{\hat x}_{i-1}$ by a mixer network, formed by two sequential convolutional layers, to produce the final reconstruction $\bm{\hat x}_{i}$.

Results of this scheme can be seen in Fig.~\ref{fig:resnet-awgn} for the same scenario considered in Fig.~\ref{fig:single2}, using $m=10$ for the received image estimations. The results show that the scheme is able to achieve results very close to the previous schemes. As expected, the first layer performance is exactly the same as single transmission with rate $k/n = 1/12$, given that the base layer is trained without the knowledge of subsequent layers. Particularly interesting, however, is the fact that the network is able to predict a valid residual representation, from the estimation of the channel using only $m=10$ independent realizations.

The main advantage of this scheme is the fact that each encoder/decoder pair can be optimized separately, given the result of the previous layers. Although this is more computationally demanding, it allows design flexibility; as opposed to the first two architectures, this architecture allows adding new layers as required, without the need to retrain the whole encoder/decoder network from scratch.
This could be used, for example, in a dynamic system that adds refinement layers as resources become available, or in a distributed communication setting, in which relay transmitters located at different regions complement the transmission by sending refinement images.
We would like to note that this scheme can also be combined with other transmissoin techniques, i.e., a refinement layer can be transmitted for a base layer transmitted using a digital separation-based approach, in which case, the exact reconstruction at the receiver would be known.

\subsubsection{Architecture Comparison}

In the previous sections three alternative DeepJSCC-$l$ architectures for successive refinement have been introduced: (a) multiple decoder networks, (b) a single decoder network, and (c) residual transmission. Numerical results show that all three architectures achieve nearly the same performance, suggesting that they can all produce successively refinable representations of natural images over an AWGN channel.
However, while all the schemes are equivalent in terms of performance, other aspects can be considered when choosing the architecture to be used in practice.

In terms of memory complexity, the single decoder architecture has clear advantages, as it requires only one encoder and one decoder network, regardless of $L$. The residual transmission scheme is the most expensive, as for every layer in $L$, a new pair of encoder and decoder has to be trained. The multiple decoder scheme needs training different decoders per layer, but has only one encoder regardless of $L$.

In terms of time (computational) complexity, as the encoder and decoder blocks used in all the schemes are the same, all but the residual transmission scheme have equivalent complexity during inference phase. The residual transmission scheme has additional overhead, as it has the extra steps of emulating each transmission $m$ times and mixing different layers' outputs.
In terms of time complexity during training, the multiple decoder architecture has advantage over the others as it can train all the layers simultaneously and in parallel, given that each layer has its own decoder. The single decoder scheme increases the time complexity of the training, as different layers should be trained sequentially, requiring more iterations of the algorithm until convergence. Lastly, the residual transmission scheme has the highest training time complexity, as apart from having to train each layer sequentially (as in the single decoder), it also has to train the mixer component to estimate the image reconstructions at the receiver.
However, as stated previously, although more memory and time consuming during training, the residual network is the only scheme that allows the addition of new layers a \textit{posteriori}, without the need of retraining the networks of previous layers.

\section{Multiple Description JSCC}

\begin{figure}
	\begin{center}
    \resizebox {0.85\linewidth} {!}%
    {

 \tikzstyle{txt} = [text centered]
 \tikzstyle{box} = [rectangle, rounded corners, minimum width=2.5cm, minimum height=1.5cm, text width=2.5cm, text centered, draw=black]
 \tikzstyle{bbox} = [rectangle, thick, minimum width=1.5cm, minimum height=1cm, text width=1.5cm, text centered, draw=black]
 \tikzstyle{arrow} = [thick,->,>=stealth]
 \tikzstyle{fitted} = [draw=gray, thick, dotted, inner sep=0.75em]


\begin{tikzpicture}[node distance=2.3cm]

\node (x) [txt, font=\fontsize{14}{0}\selectfont] {$\bm{x}$};
\node (encoder) [box, right of=x, font=\fontsize{12}{12}\selectfont] {Encoder \\ ($f_{\bm{\theta}}$)};
\node (z) [txt, right of=encoder, xshift=0.5cm, font=\fontsize{14}{0}\selectfont] { };
\node (channel1) [bbox, above right of=z,  font=\fontsize{12}{12}\selectfont, xshift=0.5cm, yshift=1.0cm] {channel};
\node (decoder1) [box, right of=channel1, xshift=1.5cm,  font=\fontsize{12}{12}\selectfont] {Decoder $01_2$\\($g_{\bm{\phi_{01}}}$)};
\node (xhat1) [txt, right of=decoder1,font=\fontsize{14}{0}\selectfont] {$\bm{\hat{x}_{01_2}}$};

\node (channel2) [bbox, below right of=z,  font=\fontsize{12}{12}\selectfont, xshift=0.5cm, yshift=-0.5cm] {channel};
\node (decoder2) [box, right of=channel2, xshift=1.5cm,  font=\fontsize{12}{12}\selectfont] {Decoder $10_2$\\($g_{\bm{\phi_{10}}}$)};
\node (xhat2) [txt, right of=decoder2,font=\fontsize{14}{0}\selectfont] {$\bm{\hat{x}_{10_2}}$};

\node (decoder3) [box, below of=channel1, xshift=6.5cm,  font=\fontsize{12}{12}\selectfont] {Decoder $11_2$\\($g_{\bm{\phi_{11}}}$)};
\node (xhat3) [txt, right of=decoder3,font=\fontsize{14}{0}\selectfont] {$\bm{\hat{x}_{11_2}}$};

\draw [arrow] (x) -- (encoder);
\draw [arrow] (encoder.east) 
-- node[above,font=\fontsize{14}{0}\selectfont] {$\bm{z}$} ++(1, 0) |- node[above right,font=\fontsize{14}{0}\selectfont] {$\bm{z_1}$}  (channel1.west);
\draw [arrow] (channel1) -- node[above,font=\fontsize{14}{0}\selectfont] {$\hat{\bm{z}}_1$} (decoder1);
\draw [arrow] (decoder1) -- (xhat1);
\draw [arrow] (encoder.east) 
-- ++(1, 0) |- node[below right,font=\fontsize{14}{0}\selectfont] {$\bm{z_2}$}  (channel2.west);
\draw [arrow] (channel2.east) -- node[above left,font=\fontsize{14}{0}\selectfont] {$\hat{\bm{z}}_2$} (decoder2.west);
\draw [arrow] (decoder2) -- (xhat2);

\draw [arrow] (channel1.east) -- ++(0.75, 0) |- (decoder3.165);
\draw [arrow] (channel2.east) -- ++(0.75, 0) |- (decoder3.187);
\draw [arrow] (decoder3) -- (xhat3);


\end{tikzpicture}

}%
	\end{center}
  \caption{DeepJSCC-$l$ for multiple descriptions problem, where all possible subsets of channel outputs are received and decoded by decoders. Here, with $L = 2$, decoders $01_2$ and $10_2$ reconstruct the image using distinct sets of channel outputs, while decoder $11_2$ uses all available information for its reconstruction. Note that decoder indexes and parameters, and reconstructions are indexed in binary base.}
  \label{fig:modelmultdescr}
\end{figure}
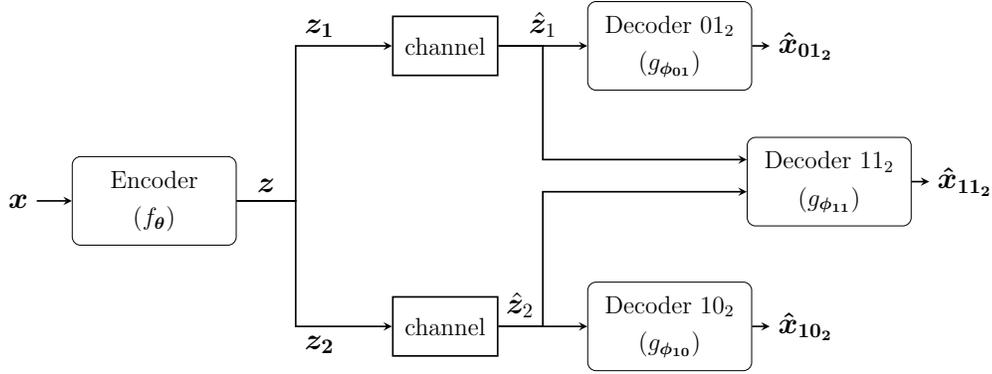

In multiple description communications we still transmit the image over $L$ parallel channels, but we have a distinct virtual decoder corresponding to any subset $\mathscr{S} \subseteq [L]$ of these channels.
For example, with $L=2$ layers, we have three potential decoders, as illustrated in Fig.~\ref{fig:modelmultdescr}. While decoders $01_2$ and $10_2$, each decodes the underlying image from only one of the layers, decoder $11_2$ decodes the same image using both layers. 
In general, all possible subsets can be indexed with binary numbers formed by $L$ bits, so that the $i$-th least significant bit is $1$ if $i \in \mathscr{S}_j$ or $0$ otherwise. Thus, we can have a total of $2^L-1$ decoders (excluding the empty subset), for all possible combinations of channel outputs.

Note that, in the $L=2$ case, if we remove Decoder $10_2$ we recover the successive refinement problem. The multiple description problem is a generalization of the successive refinement problem, and it is considerably more challenging as it has to be able to combine any subset of the channel outputs to reconstruct the image, and hence, there is no natural ordering of the transmissions into layers. In general, multiple description coding should be used when each codeword can be received independently, whereas successive refinement is more appropriate when there is an ordering among the channels, i.e., the signal over the $i$-th channel is received successfully iff all the previous transmissions are received. For example, this might be the case if the channels are ordered in time, and the receiver stops after receiving a random number of channels. On the other hand, if we consider transmission over an OFDM system with $L$ subcarriers, where different receivers are capable of receiving over different subsets of the subcarriers. The transmitter will need to employ a multiple description encoding scheme to guarantee that the image can be reconstructed by tuning into any subset of the subcarriers.

Similarly to the previous section we will present different possible architectures to realize multiple description coding; however, since the layers are independent and not sent sequentially, the residual transmission model does not apply here.

\subsection{Multiple Decoder Architecture}

The encoder-decoder DeepJSCC-$l$ architecture with a single encoder network and multiple decoders proposed in Section \ref{sec:succrefinement} can be expanded and adapted to the multiple description problem. A single encoder generates vector ${\bm z}$ with channel bandwidth $k$, and $2^L-1$ decoders are trained jointly using as inputs all different channel output subsets. Thus, we modify Eq.\  (\ref{eq:lossdirect}), producing the following loss function:

\begin{equation}
  \mathcal{L} = \frac{1}{(2^L-1)N}\sum_{j=1}^{2^L-1}\sum_{i=1}^N d(\bm x^i, \bm{\hat x}_{j}^{i}).
  \label{eq:lossmdescr}
\end{equation}

\begin{figure}
	\begin{center}
 		\subfloat[]{\includegraphics[width=0.45\textwidth]{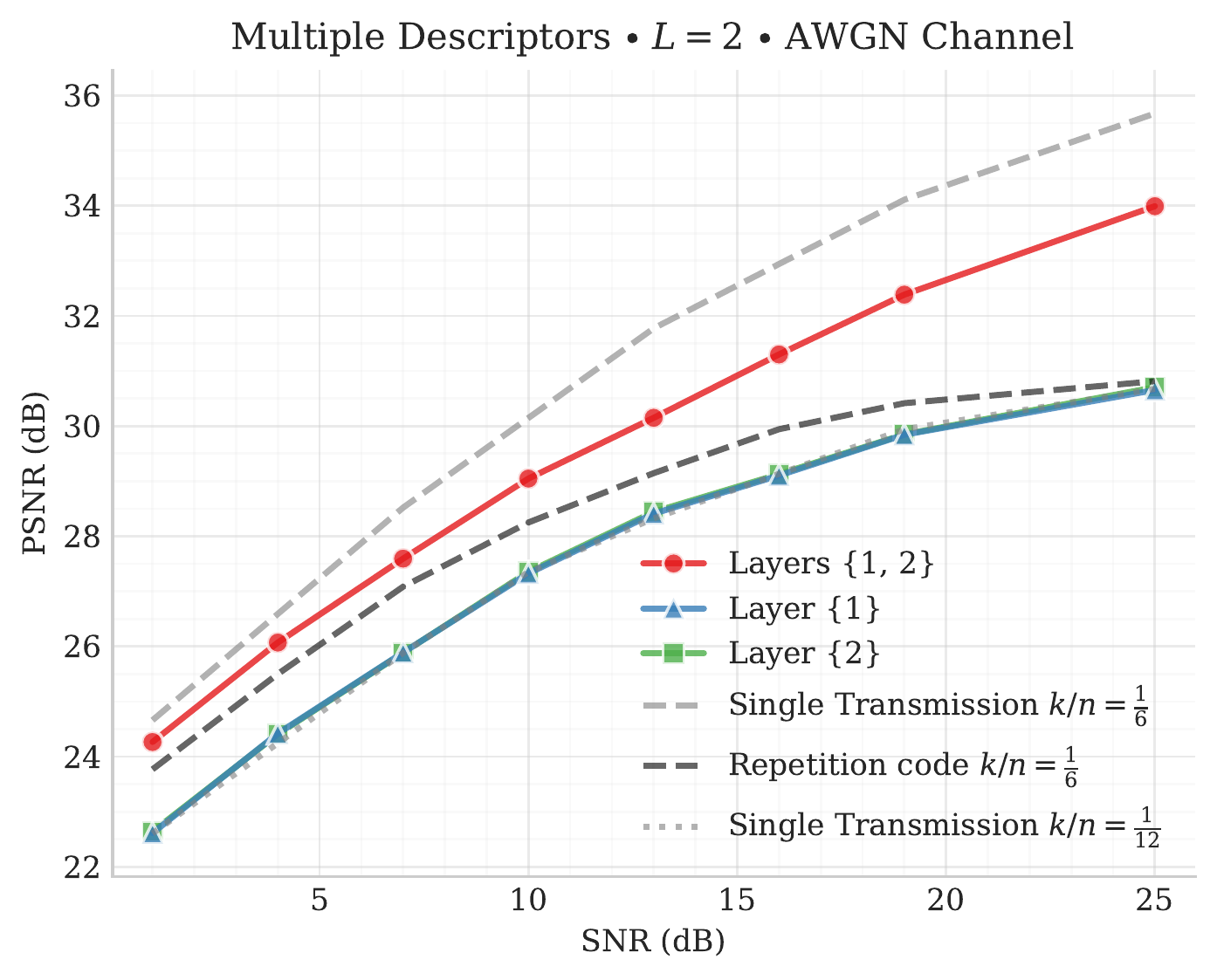} \label{fig:descr2}}
		\subfloat[]{\includegraphics[width=0.45\textwidth]{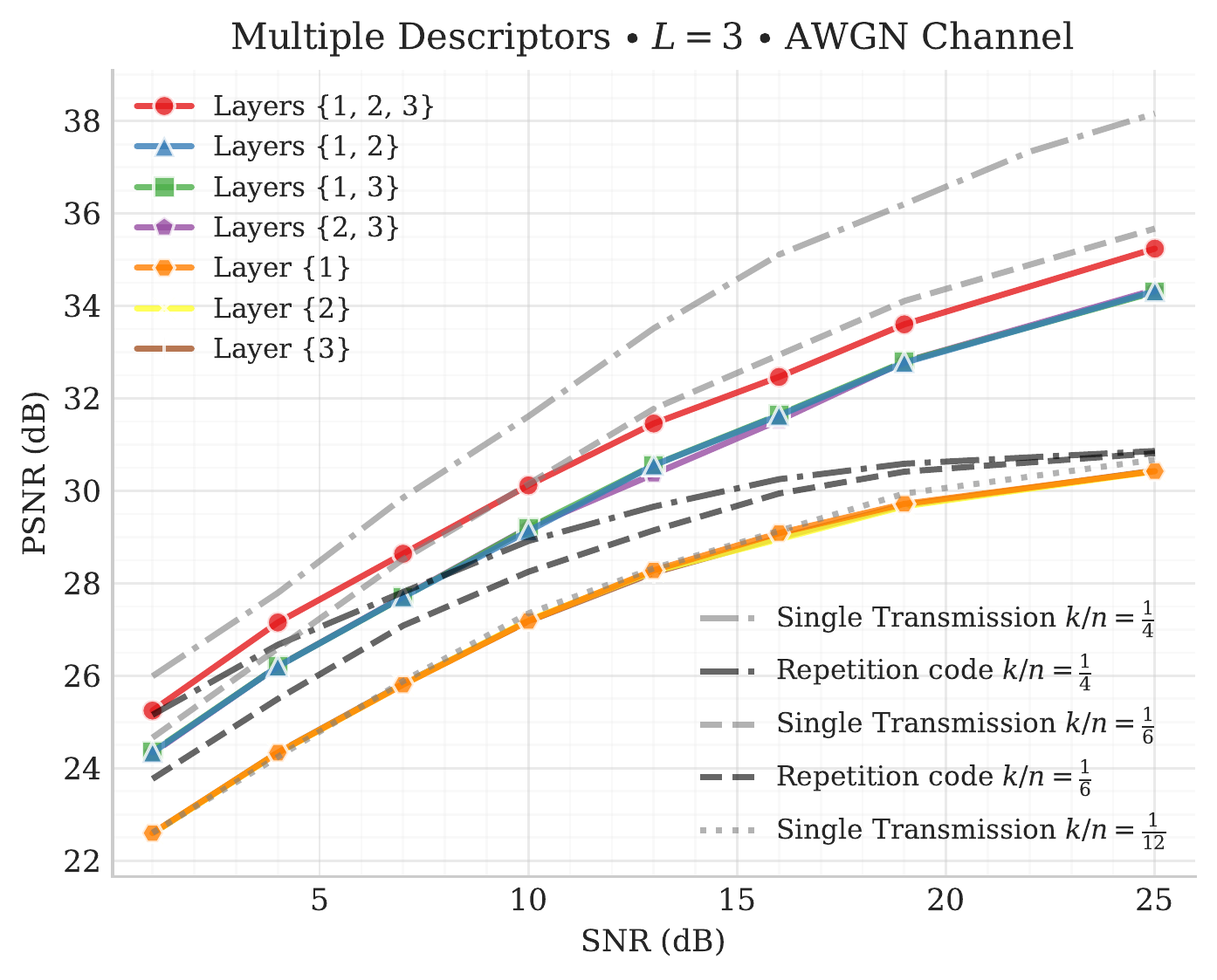}\label{fig:descr3}}
	\end{center}
		\caption{Performance of multiple description problem on CIFAR-10 test images. (a) $L=2$, AWGN channel, $k_1/n = k_2/n = 1/12$ (b) $L=3$, AWGN channel, $k_1/n = k_2/n = 1/12$.
		}\label{fig:descr}
\end{figure}

Fig.~\ref{fig:descr2} and \ref{fig:descr3} show results for $L=2$ and $L=3$, respectively. We consider individual layers with constant size (i.e., $k_i = k/L, ~\forall i \in 1, \dots , L$), so the decoders work with bandwidth as multiples of $k/L$. In all the experiments, we consider $k_i/n = 1/12$ as the bandwidth ratio. 
We see in these figures that the quality of the reconstruction of all the decoders with a single layer (i.e., $k/L$ bandwidth) is equivalent, and is almost as good as what a single layer encoder with the same dimension would produce.
When more than one layer is available, the decoder can reconstruct the input image with much better quality compared to the single-layer decoders; the combined performance, however, is inferior to a scheme which would only target the joint decoder. This is in contrast to the successive refinement problem, in which case the successive refinability could be achieved with almost no loss in the final performance. This performance loss is expected, and can be explained by the fact that, as each single-layer receiver tries to reconstruct the whole input ${\bm x}$ on its own, the information context common to both increases, and as a result, the amount of information available for the multi-layer decoders decreases. Such a rate loss is also observed in theoretical results for multiple description coding.
For example, while Gaussian sources are successively refinable; that is, they can be compressed into multiple layers, each reconstruction operating on the optimal rate-distortion curve, this is not possible in the case of multiple description coding \cite{Ozarow1980source}.
We also consider another baseline where a simple repetition code is used to transmit the same codeword over both channels, and the receiver decodes using the average of the different channel output signals it receives. Effectively, multiple codewords in this scheme are used to reduce the effect of channel noise, rather than providing complementary source descriptions. Therefore, this scheme provides gains from multiple layers only in the low SNR region, and we can see that our model outperforms this baseline at all SNR values, indicating that different layers learn complementary representations of the image, transmitting more information than a single layer repeated.

\begin{figure}
	\begin{center}
 		\subfloat[original image]{\includegraphics[width=0.5\textwidth]{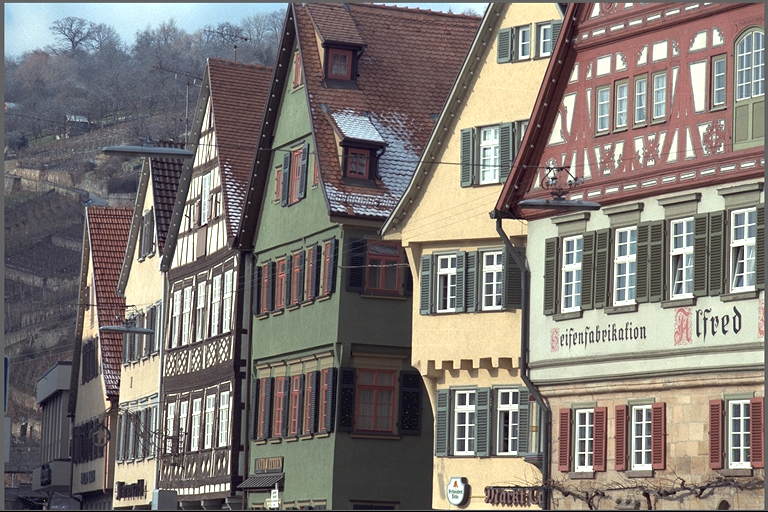}\label{fig:visual_md_original}} 
		\subfloat[$\bm{\hat{x}_{11_2}}$\protect\\PSNR: 26.85 /
		MS-SSIM: 0.9671
		]{\includegraphics[width=0.5\textwidth]{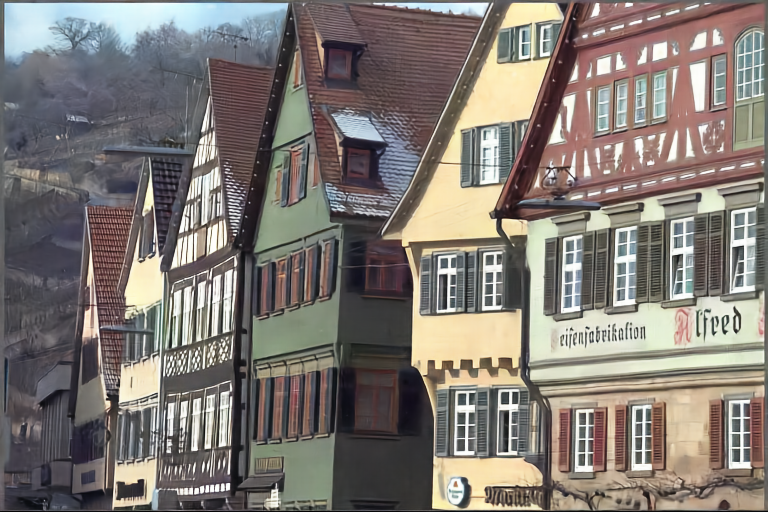}\label{fig:visual_md_layer11}}\\
		\subfloat[$\bm{\hat{x}_{01_2}}$\protect\\PSNR:  24.98 /
		MS-SSIM: 0.9487 ]{\includegraphics[width=0.5\textwidth]{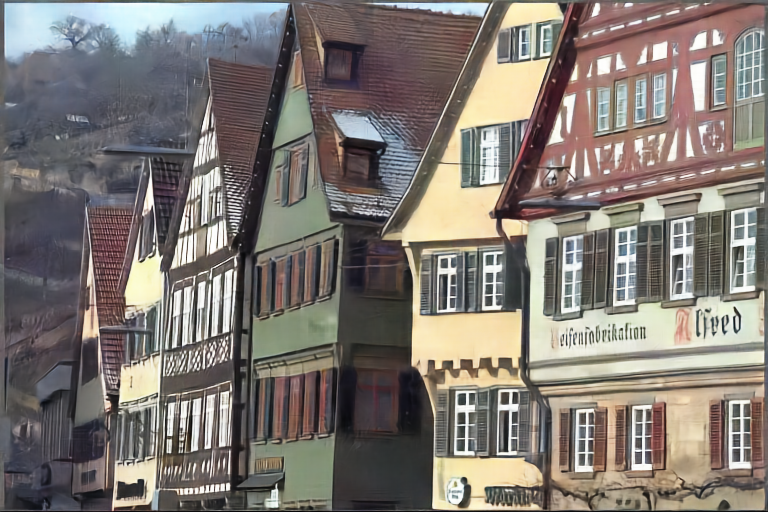}\label{fig:visual_md_layer10}}
		 \subfloat[$\bm{\hat{x}_{10_2}}$\protect\\PSNR: 24.96 /
		MS-SSIM: 0.9472]{\includegraphics[width=0.5\textwidth]{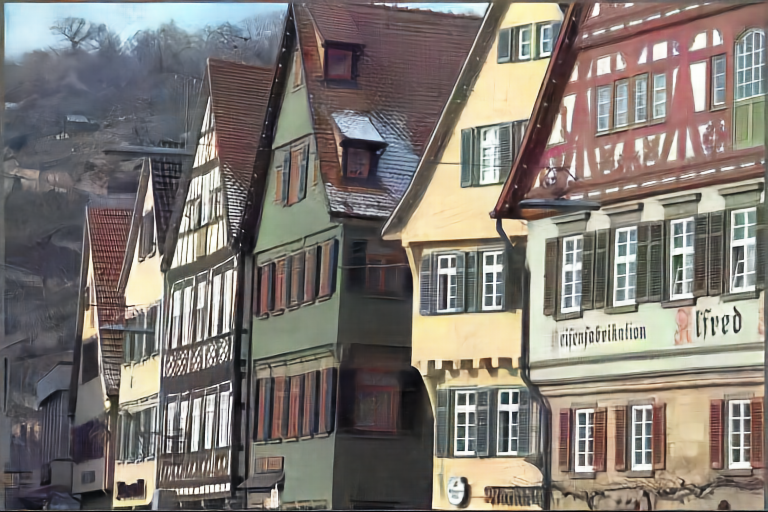}\label{fig:visual_md_layer01}} 
	\end{center}
  \caption{Examples of reconstructions for different subsets of multiple description problem for $L=2$.}\label{fig:visual-multdescr}
\end{figure}

As in Section \ref{sec:digital}, the encoder and decoder architectures of the multiple descriptor model can be extended and trained in more complex datasets. Fig.~\ref{fig:visual-multdescr} shows samples of images from the Kodak dataset from the extended model (\cite{kurka:IZS:2020}) trained on Imagenet. While the improvement from both layers is difficult to observe by naked eye at this resolution, certain details can be noticed, e.g., letter ``A'' is more clear in Fig.~\ref{fig:visual_md_layer11} compared to the other two reconstructions.

\subsection{Single Encoder-Decoder Network}

\begin{figure}
	\begin{center}
 		\subfloat[]{\includegraphics[width=0.45\textwidth]{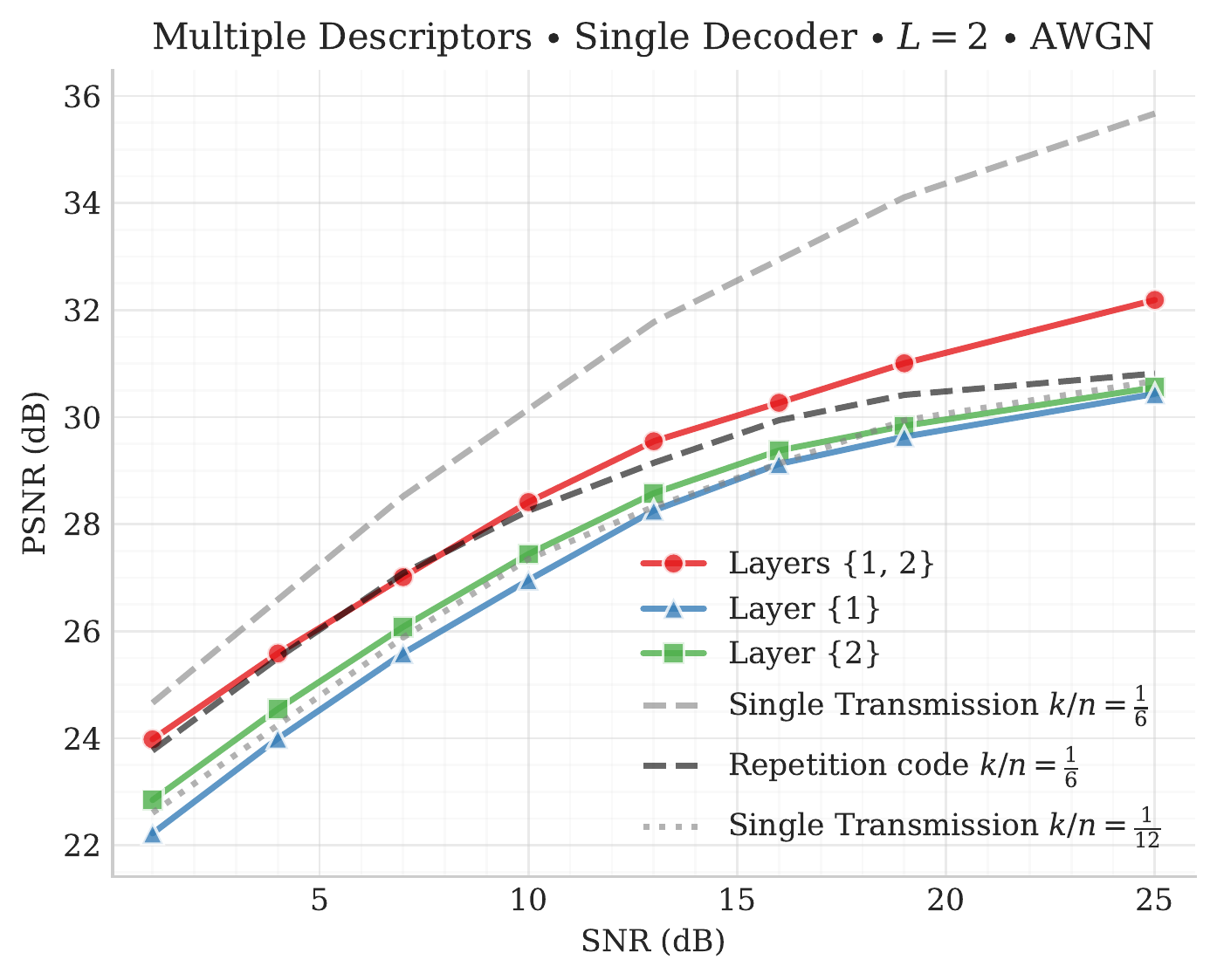} \label{fig:descrsing2}}
		\subfloat[]{\includegraphics[width=0.45\textwidth]{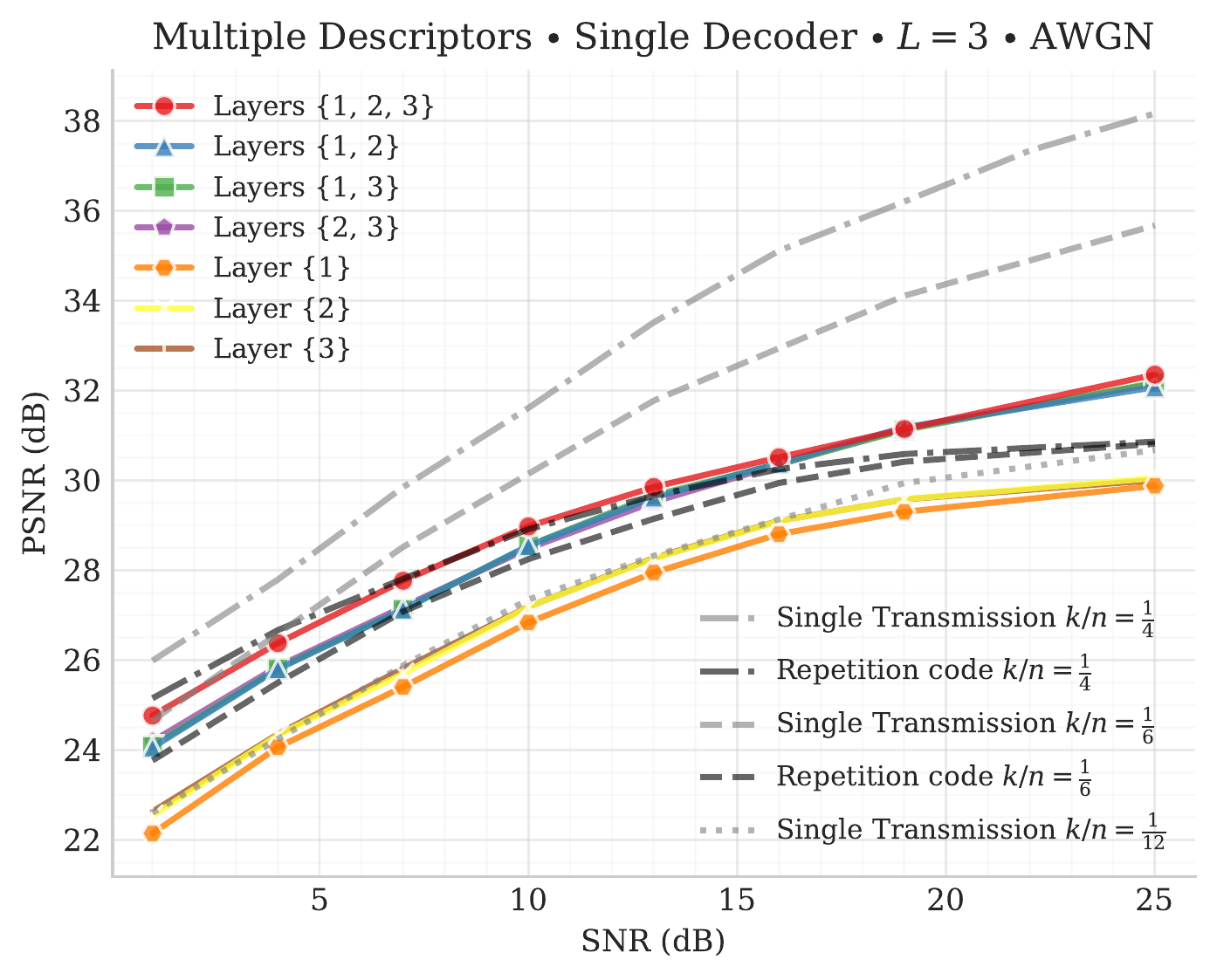}\label{fig:descrsing3}}
	\end{center}\
		\caption{Performance of the single decoder architecture, with the same configurations as in Fig.~\ref{fig:descr}.
		}\label{fig:multdescr_singledec}
\end{figure}

The single decoder model can be adapted to this scenario by simply training a decoder that inserts zeroes on blocks that are not received, according to their positions in the latent vector. The training and evaluation procedures remain the same as in the successive refinement case.
The same trade-offs apply here, that is, while multiple decoders save in training time, the single decoder saves in memory. Note, however, that the number of subsets of possible channel output layers increase exponentially with $L$, making the single decoder's learning task much more complex.
Fig.~\ref{fig:multdescr_singledec} shows results for $L=2$ and $L=3$. Although the results are not as good as those of the multiple decoders, single layer transmission has comparable performance to a single transmission with equivalent bandwidth, and transmission with multiple layers are in general better than the repetition scheme with the same bandwidth.

\section{Summary and Conclusions}

We have explored the use of deep learning based methods for progressive image transmission over wireless channels.
Building on recent results showing that artificial neural networks can be very
effective in learning end-to-end JSCC algorithms, we explored whether the
network can be extended to also learn successive refinement strategies, which
would provide additional flexibility.

We introduced DeepJSCC-$l$, a group of deep-learning based JSCC algorithms able to encode and decode images over multiple channels, allowing flexible and adaptive-bandwidth transmissions.
To the best of our knowledge, this is the first time that a hierarchical JSCC scheme has been developed and tested for practical information sources and channels.

We presented a series of architectures and experimental results, highlighting practical applications for DeepJSCC-$l$. The results show the versatility of the model to not only learn the layered representation (for both successive refinement and multiple description problems), but also comparable performance when compared to state-of-the-art methods over a wide range of SNRs and channel bandwidths. Adaptability to environmental changes is also demonstrated, with the model showing graceful degradation when there is mismatch between the design and the deployment channel qualities, and the possibility to learn to operate in diverse channels, such as fading channels.

\appendices

\section{Multi-Objective Trade-offs}
\label{sec:weights}

Both the successive refinement and multiple description problems are formulated as a multi-objective problem. Our models made the assumption that all objectives have equal weights, as shown in Eqns. (\ref{eq:lossdirect}) and (\ref{eq:lossmdescr}), in which all the losses are averaged with equal weights. Alternative approaches can also be considered with different weights.

\subsection{Successive Refinement Trade-offs}
In the successive refinement problem, one can consider that each layer's reconstruction has different weights, so reconstructions with less or more bandwidth can be prioritized.
Thus, we can rewrite Eqn. (\ref{eq:lossdirect}) as:

\begin{equation}
  \mathcal{L} = \frac{1}{L}\sum_{j=1}^L \lambda_j d(\bm x^i, \bm{\hat x}_{j}^{i}),
  \label{eq:lossdirectweights}
\end{equation}
where $\sum_{j=1}^L \lambda_j = 1$.

\begin{figure}
    \centering
    \includegraphics[width=0.5\linewidth]{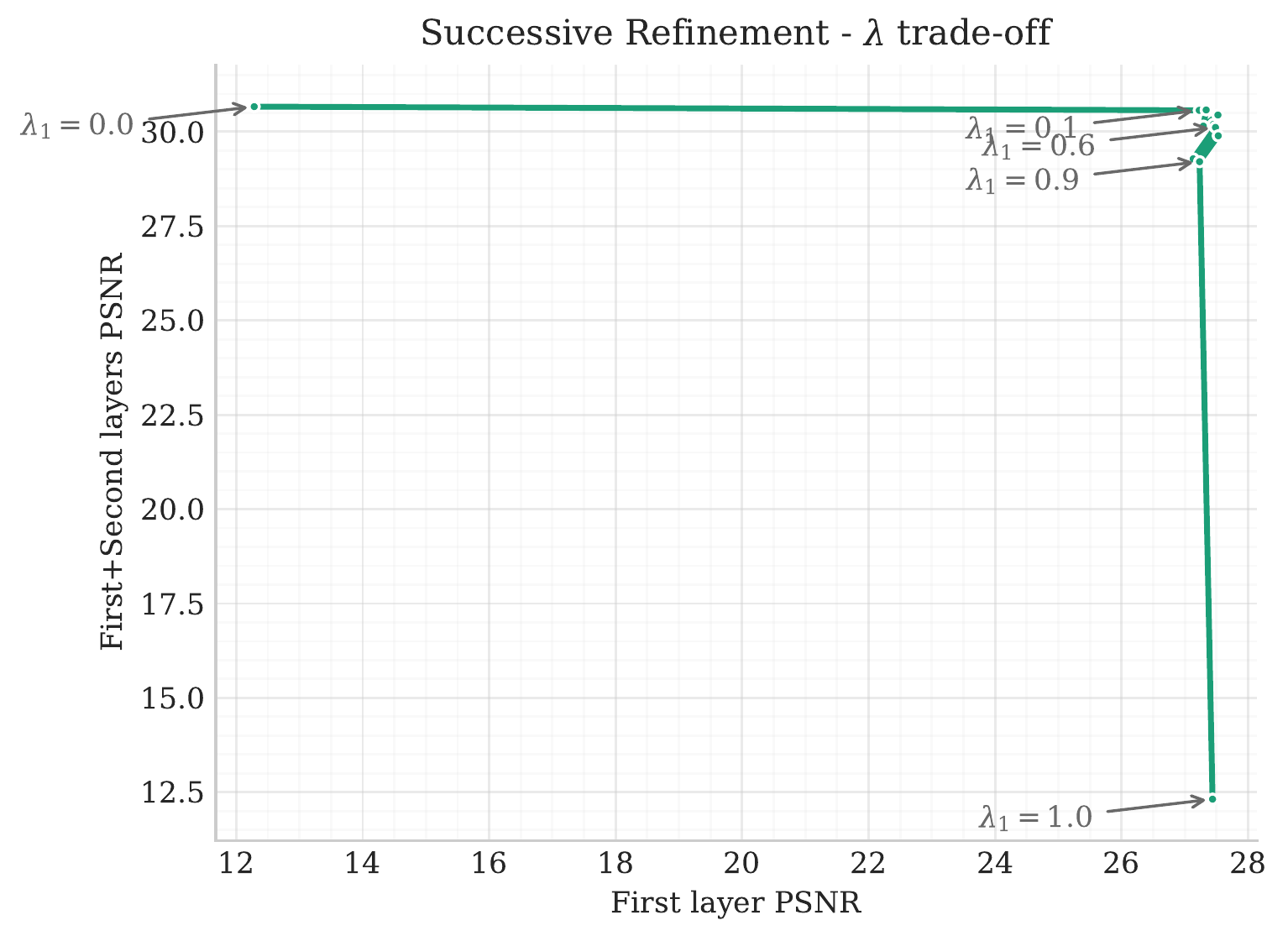}
    \caption{Trade-off between the PSNR achieved by the base layer and that is achieved by combining both layers in the successive refinement problem.}
    \label{fig:sucref_tradeoff}
\end{figure}

We consider $L=2$, and set $\lambda_1 = 1 - \lambda_2$. Fig.~\ref{fig:sucref_tradeoff} presents the simulation results. When extreme cases are considered ($\lambda_1 \cong 0$, or $\lambda_1 \cong 1$), only one of the layers dominate, as expected, with the performance of the other diminishing ($~12.5$ dB). However, for all the other intermediate values of $\lambda_1$, the choice has small impact on the overall performance of the model. This is in line with the claim that DeepJSCC-$l$ is essentially successively refinable, so the addition of weights will not interfere in the overall performance. Therefore, we use the same weights (i.e., $\lambda_j = 1/L, \forall j \in 1, \dots, L$) in all the experiments presented in the paper.

\subsection{Multiple Description Trade-offs}

As with the successive refinement problem, a multiple description transmission scheme needs to balance multiple objectives, each corresponding to the reconstruction quality of a different subset of layers. We can simplify the trade-off between different subsets by targeting the same quality if the image is decoded from the same number of layers. We will simplify further, and assume that we only consider decoders that receive single layers (indexed by $j = 2^l, \forall l \in 1, \dots, L-1$) and the decoder that receives all the layers ($j = 2^L-1$). We will then have two different quality targets, one achieved by decoding a single layer, and the other by jointly decoding all the layers. To understand the trade-off between the two, we modify the loss function in Eqn.~(\ref{eq:lossmdescr}) adding a weight $\alpha_1$ as follows:

\begin{equation}
  \mathcal{L} = \frac{1}{N} \sum_{i=1}^N \left ( \alpha_1  d(\bm x^i, \bm{\hat x}^i_{2^L-1}) +  (1-\alpha_1) \frac{1}{L}\sum_{l=0}^{L-1} d(\bm x^i, \bm{\hat x}_{2^l}^{i}) \right ).
  \label{eq:lossmdescrto}
\end{equation}

Note that, when $\alpha_1 = 1$, we only care about the joint decoder and recover the non-layered DeepJSCC scheme, and when $\alpha_1 = 0$, we only care about the single-layer decoder, which correspond to $L$ different transmissions with $1/L$\textit{th} of the bandwidth ratio.

\begin{figure}[t]
	\begin{center}
 		\subfloat[]{\includegraphics[width=0.45\textwidth]{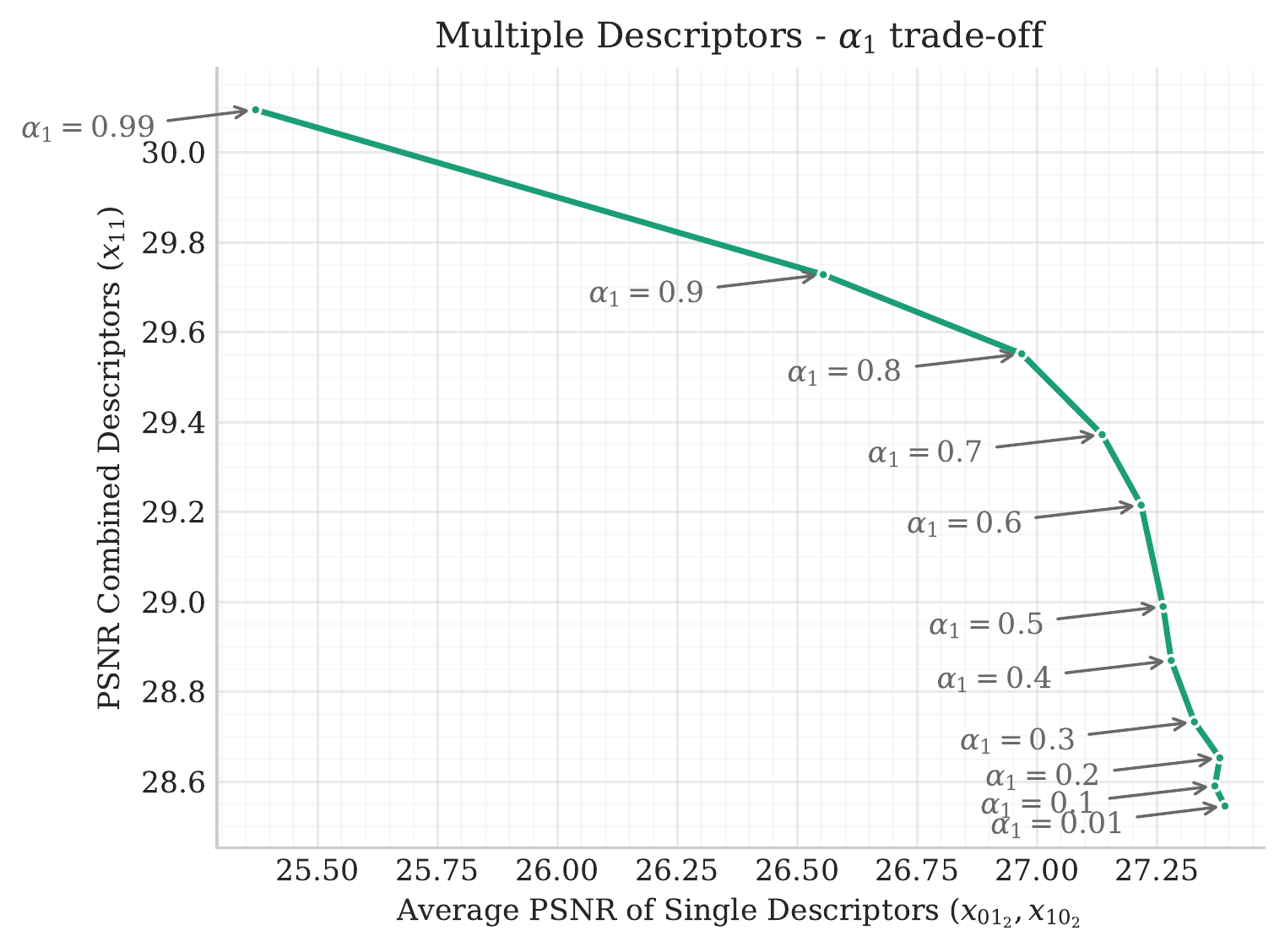} \label{fig:alpha1}}
		\subfloat[]{\includegraphics[width=0.45\textwidth]{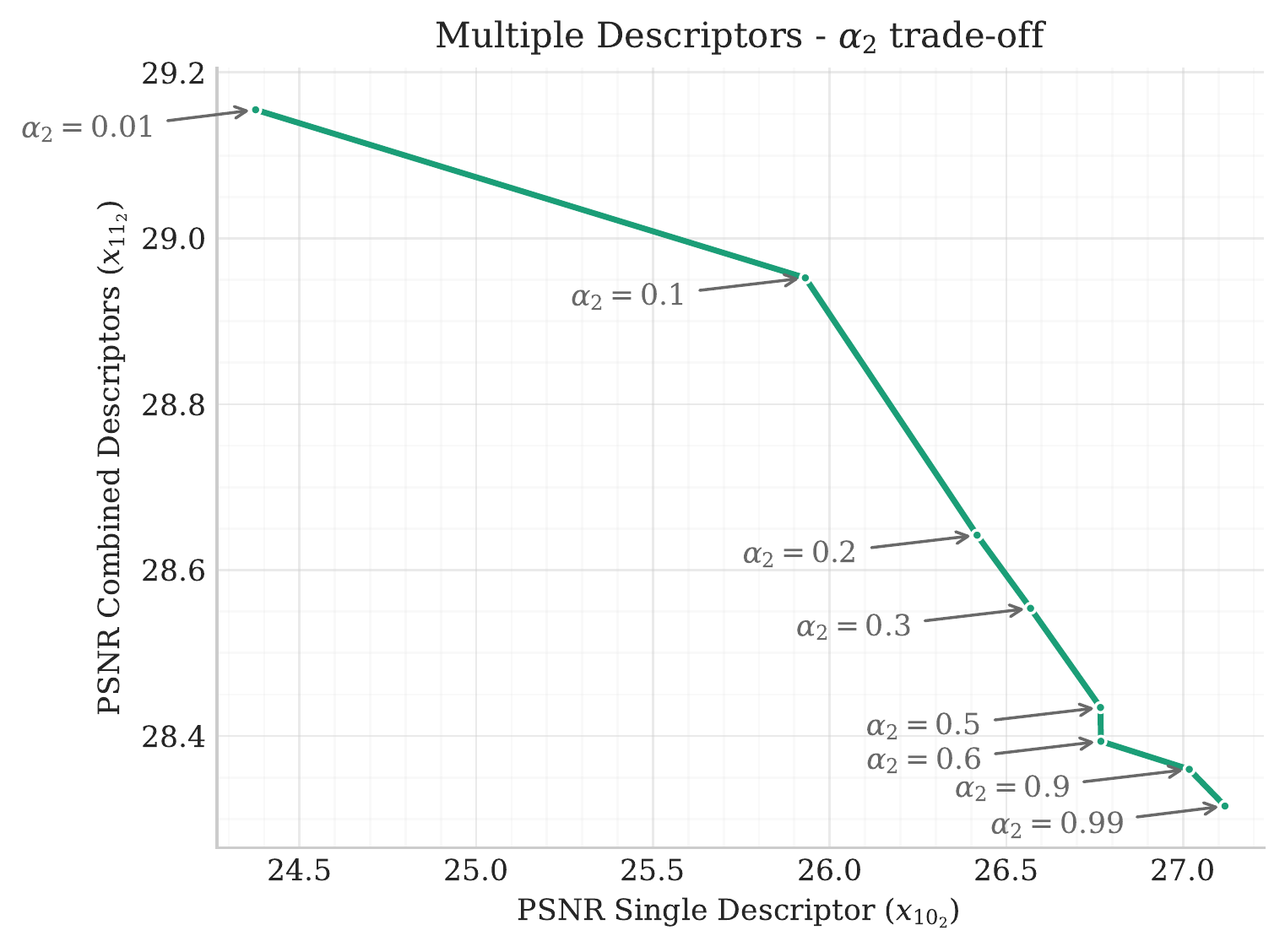}\label{fig:alpha2}}
	\end{center}
		
		\caption{Performance impact of varying the weights of different components in the multiple description problem. (a) Combined transmission vs. single components; (b) multiple description vs successive refinement.}
\end{figure}

Fig.~\ref{fig:alpha1} shows the results comparing the performance of the joint multi-layer transmission (y axis) and the average performance of single descriptor (x axis) for different values of $\alpha_1$ and $L=2$. The figure clearly illustrates the trade-off between the performance of the side and joint decoders: for small values of $\alpha_1$ the side decoders' average performance improves, approaching that of a single transmission line, as shown in Fig.~\ref{fig:descr2}. On the other hand, as $\alpha_1$ increases, the performance of the joint decoder improves, at the expense of the side decoders. When $\alpha_1$ approaches $1$, we approach the performance of a single decoder using all the available channel bandwidth.

Another possible trade-off is the choice between giving all the subsets the same weights, or prioritizing a sequence of subsets that produce successive refinement. Thus, the loss function, for the case of $L=2$ becomes:

\begin{equation}
  \mathcal{L} = \frac{1}{N} \sum_{i=1}^N \left [ (1-\alpha_2) \left(  d(\bm x^i, \bm{\hat x}^i_{11_2}) + d(\bm x^i, \bm{\hat x}^i_{01_2}) \right) + (\alpha_2) d(\bm x^i, \bm{\hat x}^i_{10_2}) \right ].
  \label{eq:multdescra1}
\end{equation}

Fig.~\ref{fig:alpha2} presents the results for different values of $\alpha_2$, comparing the performance of the second descriptor, $d(\bm x^i, \bm{\hat x}^i_{10_2})$, and the combined successive refinement transmission, $d(\bm x^i, \bm{\hat x}^i_{11_2})$. The results show the impact in the performance of the successive refinement when the second descriptor is used to independently represent a full image (instead of just refining the first descriptor). The higher the $\alpha_2$, the more emphasis is given to the decoding performance of the second descriptor alone, which decreases the performance of both descriptors combined. Finding the right balance might depend on the application and the likelihood of different subsets being experienced in the specific scenario under consideration.

\bibliographystyle{IEEEtran.bst}
\bibliography{layered.bib}

\end{document}